\documentclass[fleqn,usenatbib]{mnras}
\pdfoutput=1

\usepackage{newtxtext,newtxmath}
\usepackage{longtable,lscape}

\usepackage[T1]{fontenc}
\usepackage{ae,aecompl}

\usepackage{graphicx}	
\usepackage{amsmath}	
\usepackage{amssymb}	

\usepackage{etoolbox}
\makeatletter
\patchcmd\@combinedblfloats{\box\@outputbox}{\unvbox\@outputbox}{}{%
   \errmessage{\noexpand\@combinedblfloats could not be patched}%
}%
\makeatother

\title[Tadpole metamorphasis]{Illuminating the Tadpole's metamorphosis I: MUSE observations of a small globule in a sea of ionizing photons}

\author[M. Reiter et al.]{
Megan Reiter,$^{1}$\thanks{E-mail: megan.reiter@stfc.ac.uk (MR)}
Anna F. McLeod,$^{2,3}$
Pamela D. Klaassen,$^{1}$
Andr{\'e}s E. Guzm{\'a}n,$^{4}$
\newauthor J. E. Dale,$^{5}$
Joseph C. Mottram,$^{6}$
and Guido Garay$^{7}$
\\
$^{1}$UK Astronomy Technology Centre, ROE, Blackford Hill, Edinburgh, EH9 3HJ, UK \\
$^{2}$Department of Astronomy, University of California Berkeley, Berkeley, CA 94720, USA\\
$^{3}$Department of Physics and Astronomy, Texas Tech University, PO Box 41051, Lubbock, TX 79409, USA\\
$^{4}$National Astronomical Observatory of Japan, 2-21-1 Osawa, Mitaka, Tokyo 181-8588, Japan\\
$^{5}$Centre for Astrophysics Research, University of Hertfordshire, College Lane, Hatfield, AL10 9AB, UK\\
$^{6}$Max Planck Institute for Astronomy, K\"onigstuhl 17, 69117 Heidelberg, Germany\\
$^{7}$Departamento de Astronom\'{i}a, Universidad de Chile, Camino el Observatorio 1515, Las Condes, Santiago, Chile
}

\date{Accepted XXX. Received YYY; in original form ZZZ}

\pubyear{2019}

\begin{document}
\label{firstpage}
\pagerange{\pageref{firstpage}--\pageref{lastpage}}
\maketitle

\begin{abstract}
  We present new MUSE/VLT observations of a small globule in the Carina H~{\sc ii} region that hosts the HH~900 jet+outflow system.
Data were obtained with the GALACSI ground-layer adaptive optics system in wide-field mode, providing spatially-resolved maps of diagnostic emission lines. These allow us to measure the variation of the physical properties in the globule and jet+outflow system.
We find high temperatures ($T_e \approx 10^4$~K), modest extinction ($A_V \approx 2.5$~mag), and modest electron densities ($n_e \approx 200$~cm$^{-3}$) in the ionized gas.
Higher excitation lines trace the ionized outflow; both the excitation and ionization in the outflow increase with distance from the opaque globule. 
In contrast, lower excitation lines that are collisionally de-excited at densities $\gtrsim 10^4$~cm$^{-3}$ trace the highly collimated protostellar jet. 
Assuming the globule is an isothermal sphere confined by the pressure of the ionization front, we compute a Bonnor-Ebert mass of $\sim 3.7$~M$_{\odot}$.
This is two orders of magnitude higher than previous mass estimates, calling into question whether small globules like the Tadpole contribute to the bottom of the IMF. 
Derived globule properties are consistent with a cloud that has been and/or will be compressed by the ionization front on its surface. 
At the estimated globule photoevaporation rate of $\sim 5 \times 10^{-7}$~M$_{\odot}$~yr$^{-1}$, the globule will be completely ablated in $\sim 7$~Myr.
Stars that form in globules like the Tadpole will emerge into the H~{\sc ii} later and may help resolve some of the temporal tension between disk survival and enrichment. 
\end{abstract}




\section{Introduction}

Most stars are born in large complexes where feedback from high-mass stars will  sculpt the parent molecular cloud, and thus the natal cocoon of on-going star formation. 
Ionizing radiation carves elongated dust pillars that protrude into the H~{\sc ii} region, pointing toward the dominant ionizing source \citep[e.g.,][]{hester1996,klaassen2014,mcleod2015,mcleod2016,pattle2018}.
Many pillars have high-density clumps at their tips \citep[e.g.,][]{ohl12} that were either compressed \citep[e.g.,][]{gritschneder2010,dale2013} or uncovered by the advancing ionization front \citep[e.g.,][]{dale2011,tremblin2013}.
Initially, this high density material may cast a shadow that shields the pillar body from incident ionizing radiation.
Eventually, the clumps at the pillar tips may detach, appearing as free-floating globules in the H~{\sc ii} region.
If some globules harbor a nascent star, these may be the progenitors of the photoevaporating protoplanetary disks (proplyds) seen in nearby regions like Orion 
\citep[e.g.,][]{odell1994,mann2010,mann2014}.
Connecting these stages is critical to understanding how stellar feedback shapes ongoing star formation.

However they are formed, many H~{\sc ii} regions have a population of small, opaque globules that remain afloat in the ionized gas after the surrounding low-density material is cleared away.
The survival and star-forming potential of these oases of neutral gas have been a subject of interest for decades \citep{bok1948,pottasch1956,pottasch1958,dyson1968,herbig1974,schneps1980,reipurth1983}. 
Many theoretical investigations have explored the sensitivity of globule evolution to initial conditions and the shape and intensity of the radiation field 
\citep{sandford1982,bertoldi1989,lefloch1994,kessel-deynet2003,esquivel2007,miao2009,bisbas2011}.

Whether this external feedback triggers or starves star formation remains a matter of debate. 
Plane-parallel radiation focuses material whereas the divergent flux of a point-source results in multiple fragments \citep{esquivel2007}. 
The shock driven by the ionization front may dramatically enhance the effects of self-gravity \citep[e.g.,][]{miao2009,gritschneder2009,bisbas2011}. 
\citet{tremblin2013} argue that the observed star formation would have happened anyway, but may benefit from the accumulation of matter being pushed by feedback. 
In contrast, \citet{kessel-deynet2003} argue that the redistribution of kinetic energy into undirected random motions may prolong the cloud lifetime by making it more robust against collapse.

Globules that do collapse may make a unique contribution to the stellar and planetary populations in feedback-dominated regions.
Initially, the globule will shield the embedded protostar from ionizing radiation that rapidly destroys protoplanetary disks \citep[e.g.,][]{winter2018,nicholson2019}.
Stars forming in these globules are relatively close to the high-mass stars that provide chemical enrichment thought to play a central role in the geochemical evolution of terrestrial planets 
\citep[e.g.,][]{cameron1977,grimm1993,hester2004,lichtenberg2019}. 
As such, star-forming globules represent an intermediate case that may help reconcile short disk lifetimes with the lifetimes of high-mass stars.

Smaller globules that collapse may form planetary-mass objects, potentially contributing significantly to the low-mass end of the initial mass function \citep[IMF; see][]{grenman2014}.
If this is the case, then feedback-dominated regions may show evidence for variations in the IMF. 
While observations of local Galactic star-forming regions suggest that the IMF is invariant \citep[e.g.,][]{bastian2010}, the situation is less clear in the more extreme environments sampled in other galaxies \citep[e.g.,][]{conroy2012,conroy2013,zhang2018}. 
Local regions provide the best laboratory to resolve how feedback affects star formation and the low-mass stellar (and sub-stellar) population, unlike extra-galactic environments where direct measurements of the IMF are impossible.

\citet{grenman2014} catalogued a population of small globules in one of the nearest giant H~{\sc ii} regions, the Carina Nebula.
These so-called globulettes are larger than revealed proplyds but smaller than cannonical Bok globules \citep{bok1947,thackeray1950}. 
Based on a single wavelength intensity-tracing analysis, they estimated that most of the globulettes are roughly planetary mass.
If even a small fraction of these globulettes undergo gravitational collapse, they may contribute significantly to the low-mass end of the IMF.
However, \citet{haworth2015} showed that these low-mass globulettes are stable and require an additional perturbation to stimulate collapse. 
Estimated photoevaporation rates \citep[e.g.,][]{smith2004_finger,reiter2015_hh900} suggest that such small, low-mass objects will be rapidly ablated.
\citet{mccaughrean2002} report evidence for star formation in $\sim 15$\% of the (larger) globules seen in M16, demonstrating that at least some globules have larger mass reservoirs, and thus may survive longer.

\begin{figure*}
  \centering
$\begin{array}{cc}
    \includegraphics[trim=25mm 15mm 0mm 0mm,angle=0,scale=0.365]{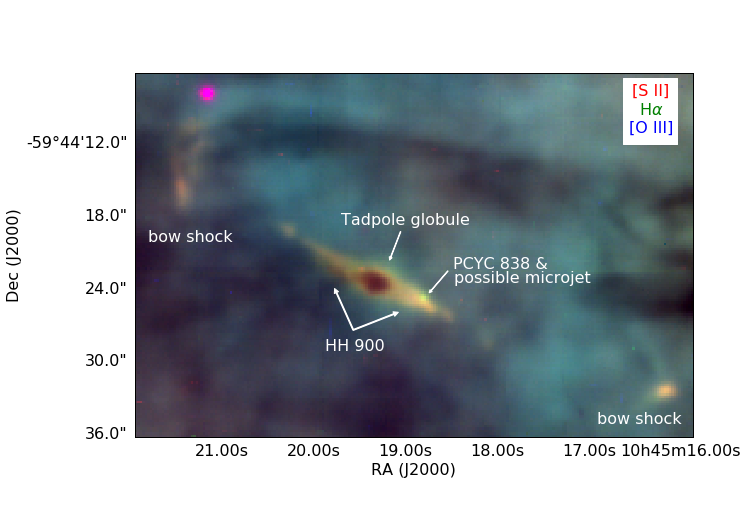}  &
    \includegraphics[trim=20mm 15mm 0mm 0mm,angle=0,scale=0.375]{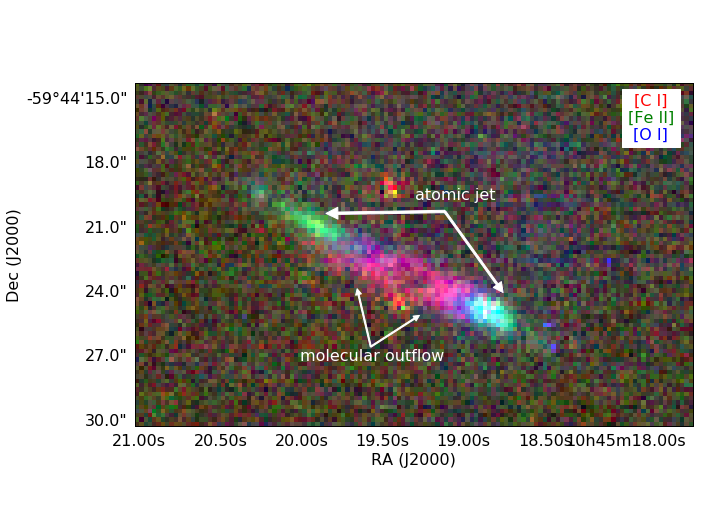} \\
\end{array}$
  \caption{
    \textit{Left:} MUSE image of the Tadpole globule and the HH~900 jet+outflow system ([S~{\sc ii}]=red, H$\alpha$=green, [O~{\sc iii}]=blue).
    A second YSO (PCYC~838) and possible microjet that lie projected onto the western limn of HH~900 are labeled. 
    \textit{Right:} Three-color image highlighting different components of the jet+outflow system: the molecular outflow ([C~{\sc i}]=red), the atomic jet ([Fe~{\sc ii}]=green), and the ionized outflow ([O~{\sc i}]=blue).
  }\label{fig:muse_intro} 
\end{figure*}

Determining the fate of small globules and constraining the relevant physics requires an estimate of the incident energy and the resulting kinematics in the cold molecular gas. 
In this paper, we consider a small \citep[$r\sim1^{\prime\prime}$, corresponding to $\sim 0.01$~pc at a distance of 2.3~kpc][]{smith2006_distance} tadpole-shaped globule in the Carina Nebula.
The so-called Tadpole (see Figure~\ref{fig:muse_intro}) lies between the young open cluster Tr16 and a large wall of gas and dust that bisects the H~{\sc ii} region. 
In optical images, the globule is seen in silhouette against the bright nebular emission from the H~{\sc ii} region.

H$\alpha$ images from the \emph{Hubble Space Telescope (HST)} revealed the unusually wide-angle outflow HH~900 \citep{smith2010} emerging from the globule and illuminated by the many O-type stars in Tr16. 
Both the morphology and the kinematics of the H$\alpha$ suggest it traces an outflow entrained by the fast, collimated jet seen in near-IR [Fe~{\sc ii}] \citep{reiter2015_hh900}. 
Protostellar jets seen in other H~{\sc ii} regions often appear to point away from the ionizing sources, bent either by their winds, the rocket effect, or both \citep[e.g.,][]{bally2006}.
However, HH~900 bends toward Tr16, leading \citet{reiter2015_hh900} to suggest that it is pushed by the photoevaporative flow off the dust wall behind it. 
With a jet dynamical age of $\sim 2200$~yr, HH~900 is one of the youngest outflows in Carina. 
Jet kinematics require a driving source embedded in the small opaque globule, but previous observations provided no evidence for a protostar inside the Tadpole.

In this paper (Paper~I), we present new data from the Very Large Telescope (VLT) using the Multi Unit Spectroscopic Explorer \citep[MUSE;][]{bacon2010} with GALACSI wide-field adaptive optics correction of the Tadpole globule and HH~900 jet$+$outflow system.
These data provide a suite of spectral lines to measure the physical properties of the hot gas in this small globule adrift in an H~{\sc ii} region.
Together with resolved observations of the cold, molecular gas in the globule (Paper~II; Reiter et al.\ in prep), 
this allows us to constrain the role of external photoionization and internal disruption of the globule by the jet, laying the groundwork for building an evolutionary scenario of feedback carved structures from pillars to globules to photoevaporating protoplanetary disks.

\section{Observations}\label{s:obs}

We obtained data with the MUSE visual wavelength panoramic integral-field spectrograph on the VLT in service mode on 03 April 2018.
Using the newly-commissioned MUSE+GALACSI Adaptive Optics module in Wide Field Mode (WFM), this provides sub-arcsec resolution over the entire $1' \times 1'$ field-of-view. 
The AO-assisted angular resolution of the MUSE observations is $\sim 0.8''$, corresponding to $\sim 0.008$~pc at the distance of Carina. 
MUSE provides a spectral resolution of R=2000-4000 over a nominal wavelength range from $4650-9300$~\AA\ with a gap between $\sim 5800-5950$~\AA\ due to the laser guide stars for the adaptive optics system. 
This corresponds to a velocity resolution of $\sim 75-150$~km~s$^{-1}$.

The Tadpole and full extent of the HH~900 outflow fit within a single MUSE pointing that we observed with a three-point dither pattern, rotating $90^{\circ}$ between each exposure to reduce instrument artifacts. 
The integration time for each exposure was 240~s, for a total on-source integration time of $720$~s. 
Data were reduced in the ESO {\sc esorex} environment of the MUSE pipeline \citep{weilbacher2012} using standard calibrations.

Due to lunar contamination, the MUSE spectra showed increased fluxes towards the blue regime of the wavelength coverage with respect to the red. This would result in falsely high fluxes of integrated maps of emission lines in that portion of the spectrum (e.g., H$\beta$, He~I $\lambda$4921, [O~{\sc iii}]$\lambda\lambda$4959,5007) leading to erroneous emission line ratios. We correct this by performing a pixel-by-pixel continuum fit and subtraction, which removes both the lunar and stellar continua simultaneously. We use the resulting continuum-subtracted cube to perform the nebular analyses discussed in this paper.

\section{Results}\label{s:results}
We detect $\sim 50$ emission lines with MUSE, with more than a dozen hydrogen recombination lines and forbidden emission lines from at least seven different atoms. 
A full list of lines detected with MUSE is presented in Table~\ref{t:MUSE_lines}. 
Detected emission lines represent multiple ionization states in the gas, although the majority have ionization potentials $\lesssim 13.6$~eV (see Table~\ref{t:ionization_potentials} for a list of first and second ionization potentials).  
The tadpole-shaped globule itself is generally seen in silhouette against the bright nebular background.
Emission lines like H$\alpha$, [N~{\sc ii}], and [S~{\sc ii}] trace the ionization front on the globule surface.
With these spatially resolved MUSE observations of diagnostic emission lines, we measure the variation in the physical conditions in the ionization front both in the globule and the externally irradiated HH~900 jet+outflow system.

Many of the emission lines that are bright in the globule ionization front also trace the wide-angle outflow originally discovered in narrowband H$\alpha$ images from \emph{HST}.
Forbidden emission lines that are collisionally deexcited in high-density gas like [Fe~{\sc ii}], [Ca~{\sc ii}], and [Ni~{\sc ii}] trace the collimated bipolar jet that threads through the center of the wide-angle outflow \citep[see Figure~\ref{fig:muse_intro} and][]{reiter2015_hh900}. 
The spectral resolution of MUSE allows us to resolve velocity differences $\gtrsim 50$~km~s$^{-1}$, enabling a kinematic analysis of the faster portions of the jet+outflow system. 
In the following sections, we derive the physical properties of the tadpole-shaped globule and the HH~900 jet+outflow system.

\begin{table}
\caption[Ionization potentials]{Ionization potentials for detected elements} 
\vspace{5pt}
\centering
\vspace{3pt}
\begin{footnotesize}
\begin{tabular}{rrr}
\hline\hline
Name & first & second \\ 
     &  (eV) &  (eV)  \\ 
\hline 
H  & 13.6 & --- \\
He & 24.6 & 54.4 \\
C  & 11.3 & 24.4 \\
N  & 14.5 & 29.6 \\
O  & 13.6 & 35.1 \\
S  & 10.4 & 23.3 \\
Cl & 13.0 & 23.8 \\
Ar & 15.8 & 27.6 \\
Ca &  6.1 & 11.9 \\
Fe &  7.9 & 16.2 \\
Ni &  7.6 & 18.2 \\
\hline
\end{tabular} 
\end{footnotesize}
\label{t:ionization_potentials}
\end{table}


\subsection{Extinction / Reddening}\label{ss:reddening}

We use the Balmer decrement to estimate the line-of-sight extinction as 
\begin{equation}
E(B-V) = \frac{E(H\beta - H\alpha)}{\kappa(H\beta) - \kappa(H\alpha)} = \frac{-2.5}{\kappa(H\beta) - \kappa(H\alpha)} \times \log_{10} \left[ \frac{(H\alpha/H\beta)_{int}}{(H\alpha/H\beta)_{obs}} \right]
\end{equation}\label{eq:general_ebv} 
where $\kappa(\lambda)$ is the value of the extinction curve at a given wavelength \citep[see the Appendix of][]{momcheva2013}. 
Assuming typical Milky Way extinction, $\kappa(H\beta) - \kappa(H\alpha) = 1.257$ \citep{fitzpatrick1999}, this becomes
\begin{equation}
E(B-V) = -1.99 \times \log_{10} \left[ \frac{2.86}{(H\alpha/H\beta)_{obs}} \right] 
\end{equation}\label{eq:ebv}
where we have assumed Case~B recombination for gas with $T = 10^4$~K and $n_e=10^2$~cm$^{-3}$ for which we expect an intrinsic flux ratio $(H\alpha/H\beta)_{int}=2.86$ \citep{osterbrock2006}. 
The observed flux ratio in and around the Tadpole globule and outflow are $(H\alpha/H\beta)_{obs} \approx 4.6-6.4$, corresponding to $E(B-V) \approx 0.4 - 0.7$.
To convert the color excess to an extinction, we adopt $R_V = 4.4 \pm 0.2$ \citep{hur2012}, which gives $2.4 \lesssim A_V \lesssim 2.8$~mag, with a median $A_V \approx 2.5$~mag.
The estimated extinction is slightly higher if we use $R_V = 4.8$, as other authors have found for Carina \citep[e.g.,][]{smith1987,smith2002}.

The extinction we estimate using the Balmer decrement is $\sim 2$~mag lower than that estimated by \citet{reiter2015_hh900} using the ratio of near-IR [Fe~{\sc ii}] lines ($A_V \approx 4.5$~mag).
Hydrogen recombination lines like H$\alpha$ trace a notably different morphological and kinematic component than the forbidden near-IR [Fe~{\sc ii}] lines in the HH~900 jet+outflow system. 
Differences in the estimated extinction may therefore reflect real differences in the extinction to different components of the system. 
More material may obscure the jet embedded within the outflow, especially if the outflow is not purely ionized gas (see Section~\ref{sss:ionized_outflow}), leading to a higher extinction to the collimated jet.
\citet{raga2015} also find that $E(B-V)$ estimates from the Balmer decrement are lower than those obtained with other lines that do not require assuming a recombination cascade. 
In the following analysis, we adopt the extinction estimated from the Balmer decrement and use \textsc{Pyneb} \citep{luridiana2015} to correct the observed fluxes.

\begin{figure*}
  \centering
  $\begin{array}{c}
    \includegraphics[trim=30mm 10mm 10mm 30mm,angle=0,scale=0.325]{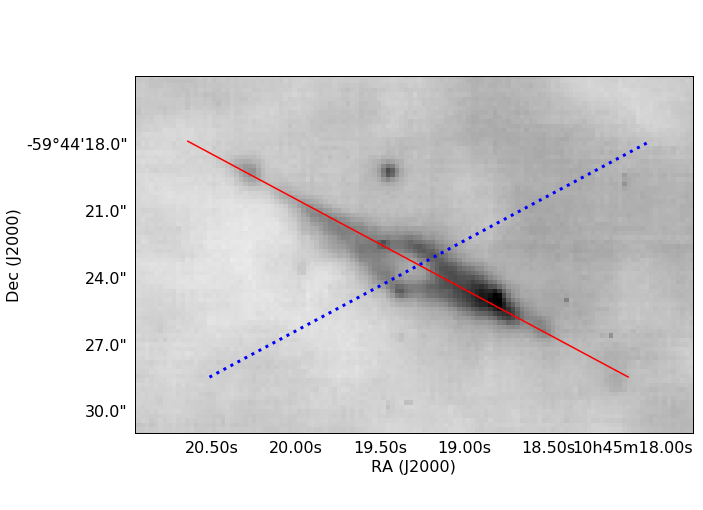} \\
    \begin{array}{ccc}
      \includegraphics[trim=0mm 0mm 0mm 0mm,angle=0,scale=0.375]{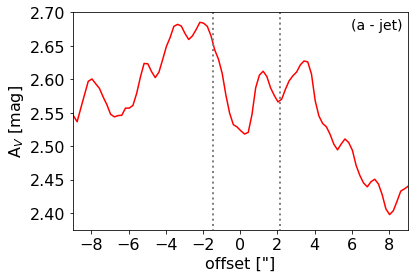} &
      \includegraphics[trim=0mm 0mm 0mm 0mm,angle=0,scale=0.25]{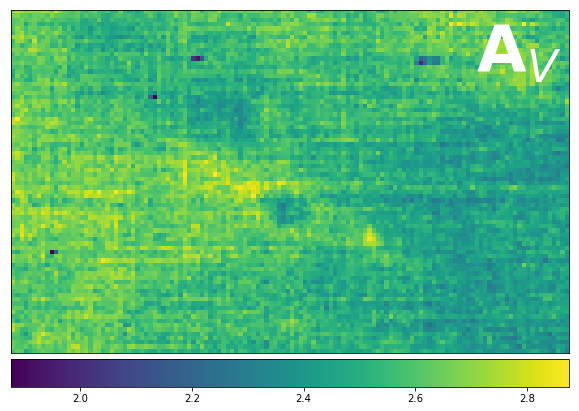} &
      \includegraphics[trim=0mm 0mm 0mm 0mm,angle=0,scale=0.375]{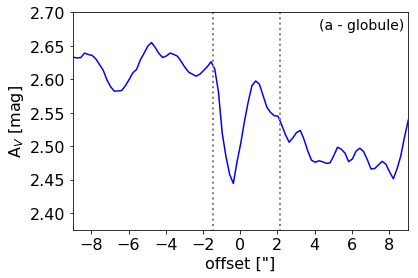} \\
      \includegraphics[trim=0mm 0mm 0mm 0mm,angle=0,scale=0.375]{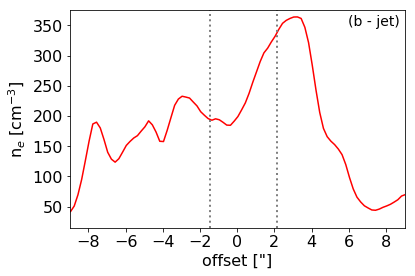} &
      \includegraphics[trim=0mm 0mm 0mm 0mm,angle=0,scale=0.25]{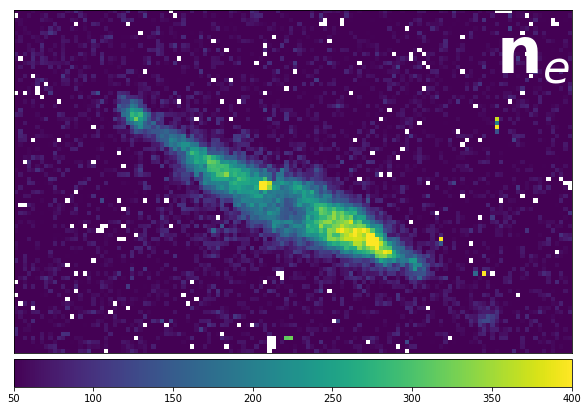} &
      \includegraphics[trim=0mm 0mm 0mm 0mm,angle=0,scale=0.375]{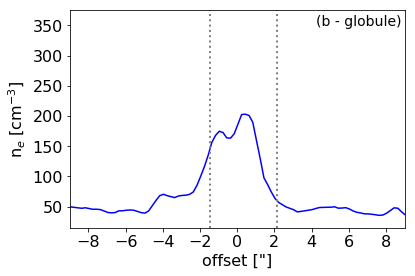} \\
      \includegraphics[trim=0mm 0mm 0mm 0mm,angle=0,scale=0.375]{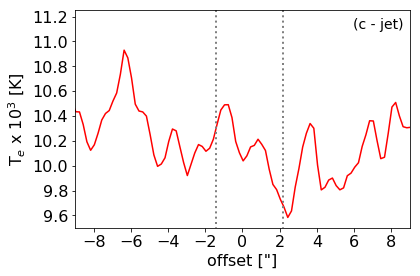} &
      \includegraphics[trim=0mm 0mm 0mm 0mm,angle=0,scale=0.25]{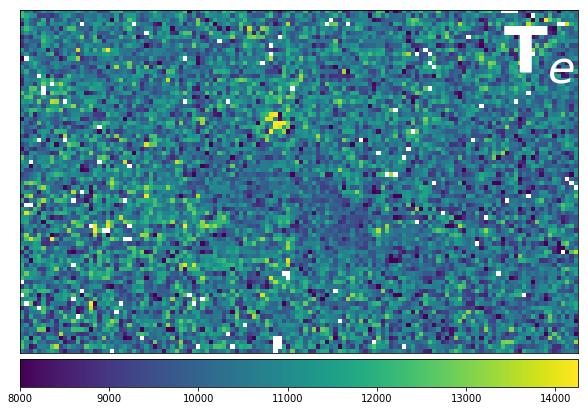} &
      \includegraphics[trim=0mm 0mm 0mm 0mm,angle=0,scale=0.375]{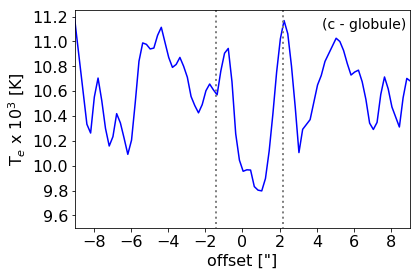} \\
      \includegraphics[trim=0mm 0mm 0mm 0mm,angle=0,scale=0.375]{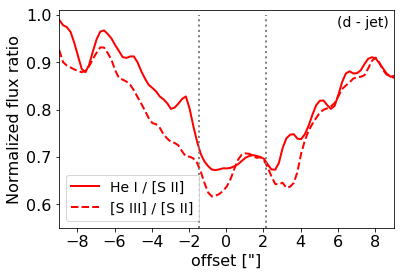} &
      \includegraphics[trim=0mm 0mm 0mm 0mm,angle=0,scale=0.25]{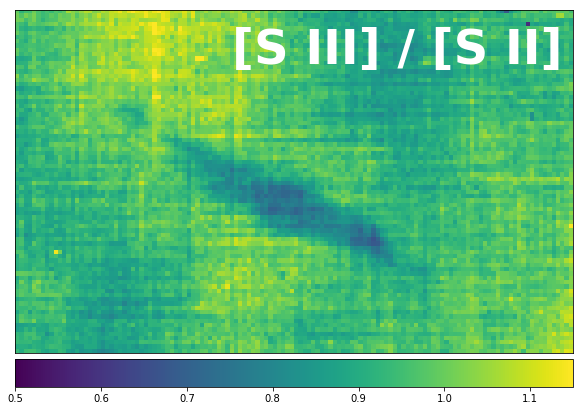} &
      \includegraphics[trim=0mm 0mm 0mm 0mm,angle=0,scale=0.375]{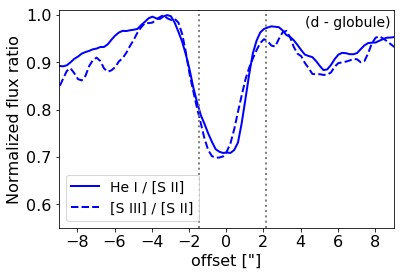} \\
\end{array}
  \end{array}$ 
  \caption{
    A [S~{\sc ii}] image showing the slices through the globule parallel and perpendicular to the jet+outflow. We plot the variation along the jet+outflow (red line; plots shown on the \textit{left}) and perpendicular to it (blue dotted line; plots shown on the \textit{right})
    of the following physical parameters (see Sections~\ref{ss:reddening} and \ref{ss:diagnostics}): 
    \textbf{(a)} $A_V$;  
    \textbf{(b)} $n_e$;  
    \textbf{(c)} $T_e$;   
    \textbf{(d)} the degree of ionization as traced by the He~{\sc i} 6678~\AA\ / ([S~{\sc ii}] $\lambda 6717 + \lambda 6731 / 2)$ ratio (black line) and 
    the [S~{\sc iii}] $\lambda 9068$ / [S~{\sc ii}] $\lambda 6717$ ratio (gray dotted line). Vertical dashed gray lines show the approximate position of the globule edges. 
  }\label{fig:physical_params} 
\end{figure*}

\subsection{Electron temperature and density in the ionized gas}\label{ss:diagnostics}

With AO-assisted IFU data, we can make spatially-resolved estimates of the density and temperature to determine how they vary as a function of position in the system. 
Many traditional nebular emission lines are detected from the wide-angle ionized outflow (both [S~{\sc ii}] and [N~{\sc ii}] have the same morphology as H$\alpha$, see Figure~\ref{fig:intensity_maps}).

Density-sensitive line ratios compare the flux of emission lines with significantly different radiative transition probabilities.
We estimate the electron density, $n_e$, from the ratio [S~{\sc ii}] $\lambda 6717 / \lambda 6731$ which is sensitive to densities in the range $10^2 - 10^4$~cm$^{-3}$. 
Emission lines that originate from significantly different excitation energy levels from the same atom provide an estimate of the temperature in the gas.
To estimate the electron temperature, $T_e$, we use the ratio [N~{\sc ii}] $\lambda6548 + \lambda6583 / \lambda5755$.
We use \textsc{pyneb} to solve for both quantities simultaneously at each pixel in the MUSE map. 
Variations in each quantity as a function of position are shown in Figure~\ref{fig:physical_params}. 
Electron densities vary from 
$35 \lesssim n_e \lesssim 370$~cm$^{-3}$ with the highest densities found in the edge of the globule and inner outflow.
The derived $T_e$ shows a smaller range of variation along the outflow / globule ($T_e \approx 10^4$~K, see Figure~\ref{fig:physical_params}), with the lowest temperatures near the globule.

Both [S~{\sc ii}] and [N~{\sc ii}] trace the wide-angle outflow, following the morphology of H$\alpha$. 
Thus the temperature and density derived from these ionized gas tracers may not reflect the physical conditions in the core of the jet seen in near-IR [Fe~{\sc ii}] emission. 
\citet{reiter2015_hh900} argue that high densities in the jet core ($\gtrsim 10^4$~cm$^{-3}$) are required to shield the Fe$^+$ from further ionization. 
\citet{nisini2002,nisini2005} also derived smaller electron densities from the [S~{\sc ii}] lines compared to those determined from the near-IR [Fe~{\sc ii}] lines. 
\citet{bautista1994} argue that the optical [Fe~{\sc ii}] lines reveal even higher densities, originating from regions with $n_e \approx 10^5 - 10^7$~cm$^{-3}$. 
We detect multiple [Fe~{\sc ii}] emission lines in the MUSE data (see Table~\ref{t:MUSE_lines}); all trace the same collimated morphology seen in the near-IR.
Using diagnostic diagrams from \citet{bautista1998}, we find that the ratio [Fe~{\sc ii}] $\lambda 7155 / \lambda 8617$ suggests $n_e > 10^4$~cm$^{-3}$, provided that photoexcitation also contributes to the emission.
Lines with lower critical densities, like the classic nebular tracers used to estimate the temperature and density of the ionized outflow, will be de-excited in regions with such high densities and are therefore a poor tracer of the physical conditions in the [Fe~{\sc ii}]-emitting jet. 
Thus, we allow that temperatures may also be lower in the high density, lower ionization jet component (see Section~\ref{ss:excitation}).

\subsection{Excitation diagnostics}\label{ss:excitation}
\begin{figure*}
  \centering
  $\begin{array}{cc}
    \includegraphics[trim=0mm 0mm 0mm 0mm,angle=0,scale=0.375]{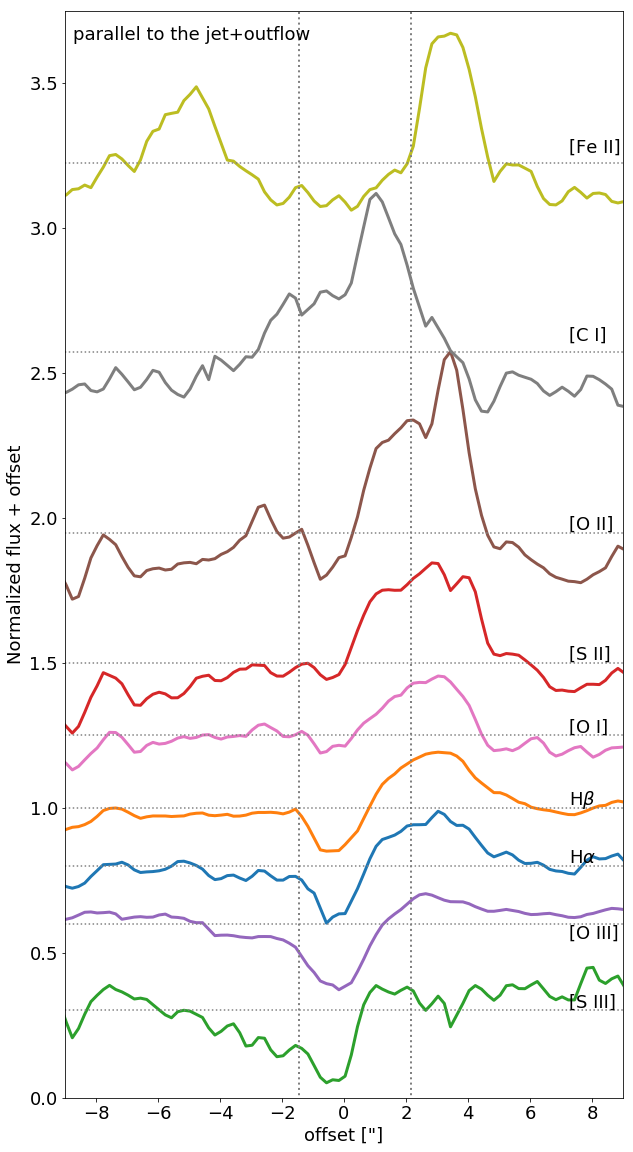} &
    \includegraphics[trim=0mm 0mm 0mm 0mm,angle=0,scale=0.375]{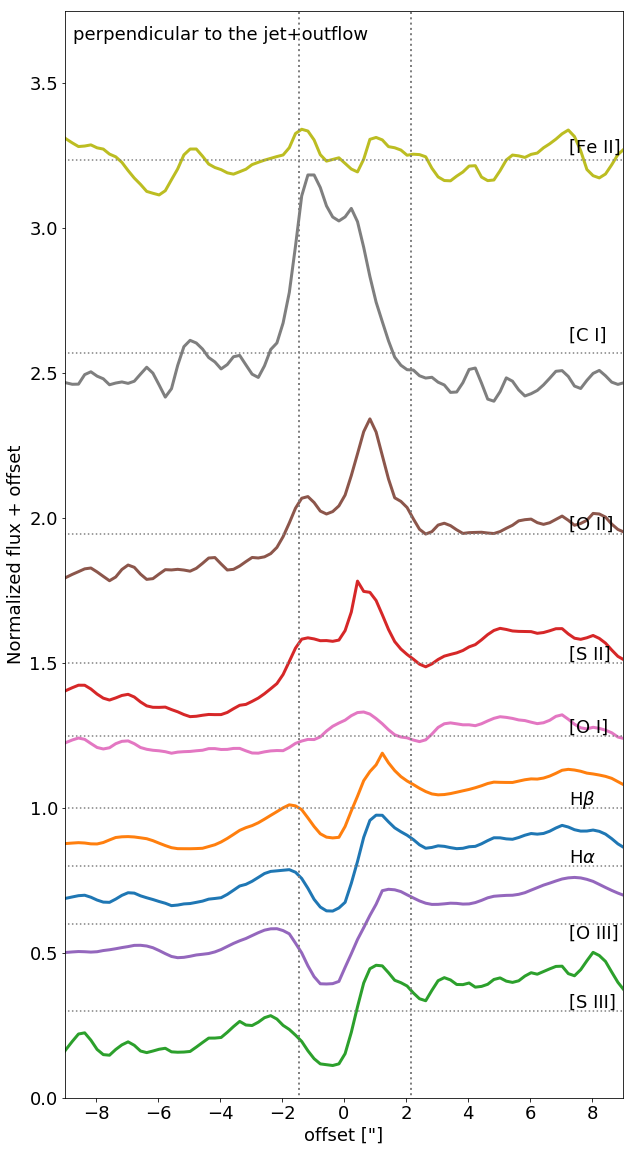} \\
  \end{array}$ 
  \caption{Emission line intensity tracings parallel (left) and perpendicular (right) to the HH~900 jet+outflow system. Cuts are made at the same locations shown in Figure~\ref{fig:physical_params}. We have added a small offset to the normalized intensity; horizontal gray dotted lines indicate the mean flux level. Vertical dashed gray lines shown the approximate edges of the globule. 
  }\label{fig:line_tracings} 
\end{figure*}

The intensity and spatial morphology of the emission lines we detect with MUSE depend on the dominant excitation mechanism(s). 
Studies of ionized interfaces often find that photoexcitation dominates over shocks \cite[e.g.][]{yeh2015,mcleod2015}.
\citet{reiter2015_hh900} argued that this is also true in the HH~900 outflow based on the H$\alpha$/[Fe~{\sc ii}] ratio, as well as the morphology and kinematics of both lines.
The spectral coverage of MUSE includes many lines that may be used in diagnostic ratios to map the excitation over the Tadpole globule and HH~900 jet+outflow system to test whether -- and where -- shock excitation dominates over photoexcitation.

Tracings of the intensity profile of several key emission lines in slices parallel and perpendicular to the outflow axis are shown in Figure~\ref{fig:line_tracings}.
For shock excitation, we expect 
[N~{\sc ii}]/H$\alpha$~$> 0.51$ and 
[S~{\sc ii}]/H$\alpha$~$> 0.45$ 
\citep{allen2008}.
However, the line ratios we measure in the Tadpole globule and the HH~900 jet+outflow system are typically much less than these values (median ratios are [N~{\sc ii}]/H$\alpha=0.12$ and [S~{\sc ii}]/H$\alpha=0.061$).
No point in the contiguous system meets or exceeds the shock-excitation threshold.
Values of the [N~{\sc ii}]/H$\alpha$ ratio are similar in the bow shocks
($\sim 0.13$ and $\sim 0.16$ for the eastern and western bow shocks, respectively), and slightly higher in the [S~{\sc ii}]/H$\alpha$ ratio ($\sim 0.7$ and $\sim 0.8$, respectively). 
Overall the line ratios indicate that photoexcitation dominates over shocks.

Multiple line ratios in this wavelength range have been used to diagnose the degree of ionization in the gas.
We take two approaches to estimate the degree of ionization:
(1) the ratio of lines of different ionization from the same element, in this case  
[S~{\sc iii}] 9069\AA\ / [S~{\sc ii}] 6717\AA; and
(2) the ratio of lines from elements with similar ionization potentials (see Table~\ref{t:ionization_potentials}), in this case 
I(He~{\sc i} $\lambda6678$) / I([S~{\sc ii}] ($\lambda6717 + \lambda6731$)/2)).
As the ionization increases, so too does the ratio of these lines since the line with the slightly lower ionization potential (the denominator) decreases as more of it gets ionized.
\citet{glushkov1998} used the latter ratio to argue that M17 has the highest degree of ionization of galactic star forming regions ($>2$ in M17 compared with $\approx 1.1$ in NGC~1976, near the Trapezium). 
Typical ratios are lower in the nebular gas surrounding the Tadpole ($\sim 0.24$) and lower still in the globule and outflow ($\sim 0.20$). 
There is a clear trend of increasing ionization with distance from the center of the globule.
In both tracers, the ratio at the terminus of the contiguous outflow is $\sim 1.5\times$ greater than the ratio measured in the least ionized gas in and immediately around the globule (see Figure~\ref{fig:physical_params}).

Most of the lines that we detect in the outflow have ionization potentials $<13.6$~eV; emission lines from more highly ionized elements requiring more energetic photons are significantly fainter. 
Notable exceptions to this trend are: 
(1) Neutral carbon, [C~{\sc i}] $\lambda 8727$, which is significantly brighter than the possible C~{\sc ii} $\lambda 6578$ line, despite the fact that the ionization potential of C is 11.3~eV. 
Both lines are seen only in the immediate vicinity of the globule.
(2) Highly ionized lines like [S~{\sc iii}] and [Ar~{\sc iii}], are seen in the entire outflow and both bow shocks.
Both require photon energies $>20$~eV to achieve the second ionization, although we do not detect lines like He$^+$ that also have ionization potentials in this range. 
(3) The detection of [N~{\sc ii}] but not [N~{\sc i}], even though the ionization potential of N is 14.534~eV.

\begin{figure}
\centering
\includegraphics[trim=10mm 0mm 0mm 0mm,angle=0,scale=0.625]{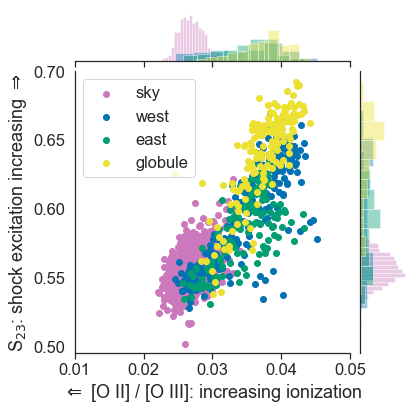} 
\caption{Same excitation / ionization parameters as explored for the nebular emission in the ONC by \citet{mcleod2016_orion}.
  All points are shown in purple, with a few select portions of the outflow (east=green, west=blue) and the globule face (gold) highlighted. 
  Ambient gas (purple points in lower left portion of the plot) tends to have higher ionization than the globule and outflow.
  The $S_{23}$ parameter shows a large spread in all components, although there is a hint that the value of $S_{23}$ is slightly larger in the western limb of the jet.
}\label{fig:ionization_parameters} 
\end{figure}
We can simultaneously compare ionization and shock excitation in the gas using the diagnostic diagram shown in Figure~\ref{fig:ionization_parameters}. 
More highly ionized gas will have smaller [O~{\sc ii}]/[O~{\sc iii}] ratios;
shock excitation will enhance the ratio $S_{23} =$ ([S~{\sc ii}] $\lambda 6717 +$ [S~{\sc ii}] $\lambda 6731 +$ [S~{\sc iii}] $\lambda 9068$)~/~H$\beta$. 
We plot the line ratio measured in individual pixels in Figure~\ref{fig:ionization_parameters} using separate colors to distinguish between different components of the system.
Points in the Tadpole system occupy a smaller portion of parameter space than sampled in the Orion Nebula Cluster \citep[the ONC, see][]{mcleod2016_orion}.
The most highly ionized gas in this study overlaps with the lower excitation proplyds in \citet{mcleod2016_orion}.
At the same time, the $S_{23}$ parameter is higher overall, with most Tadpole points lying above the values \citet{mcleod2016_orion} measure in the Bright Bar, overlapping with the lower end of the shock-excited emission measured in the HH~201 protostellar jet.
Overall, shock-excitation does not dominate and the globule and outflow are less ionized than the ambient gas surrounding the system.

\section{The irradiated HH~900 jet+outflow system}\label{ss:jet_props}

Multiple emission lines in the MUSE data trace the irradiated HH~900 jet+outflow system.
H~{\sc ii} region emission lines like H$\alpha$, [N~{\sc ii}], [S~{\sc ii}], and [O~{\sc ii}] trace the wide-angle, externally irradiated outflow.
  Line ratios in the irradiated outflow suggest that photoexcitation dominates over shocks (see Section~\ref{ss:excitation}).
In the following sections, we examine the jet and outflow components separately, and finally examine the kinematics of the jet+outflow system.

\subsection{The jet}\label{sss:jet}
  
Forbidden emission lines 
such as [Ca~{\sc ii}], [Ni~{\sc ii}], and [Fe~{\sc ii}] trace the fast, collimated jet seen in near-IR [Fe~{\sc ii}] emission \citep{reiter2015_hh900}. 
  The detection of the optical [Fe~{\sc ii}] lines suggests $n_e \gtrsim 10^5$~cm$^{-3}$ \citep{bautista1994}, indicating that densities in the jet may be an order of magnitude higher than estimated by \citet{reiter2015_hh900}. 
[Ni~{\sc ii}] traces the same jet components that are bright in [Fe~{\sc ii}], as is often observed in protostellar jets given their similar ionization potentials \citep[7.9 and 7.6 eV, respectively, see e.g.,][]{bautista1996,nisini2005}.
For $n_e \sim 10^5$~cm$^{-3}$, the [Ni~{\sc ii}] $\lambda 7412$ / $\lambda 7378$ flux ratio in the jet is not consistent with pure collisional excitation \citep{bautista1996}, suggesting signficant photoexcitation from the strong UV environment of Carina \citep[e.g.,][]{reiter2013}. 
The [Ca~{\sc ii}] $\lambda 7291 / \lambda 7324$ flux ratio is also not consistent with pure collisional excitation, with values $<1.5$ throughout the jet \citep{hartigan2004}.  
Together, this suggests that the density in the fast, collimated jet-like component of HH~900 may be an order of magnitude higher than that estimated by \citet{reiter2015_hh900}, leading to a corresponding increase in the mass-loss rate and momentum estimates.

\subsection{The ionized outflow}\label{sss:ionized_outflow}
\begin{figure*}
  \centering
  $\begin{array}{ccc}
    \includegraphics[trim=10mm 0mm 0mm 0mm,angle=0,scale=0.45]{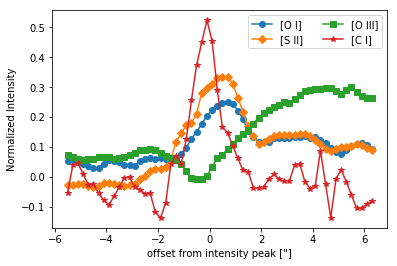} &
    \includegraphics[trim=0mm -10mm 0mm 0mm,angle=0,scale=0.475]{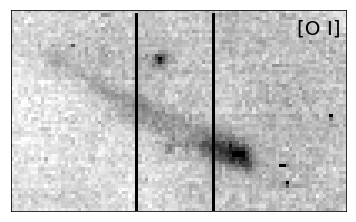} & 
    \includegraphics[trim=0mm 0mm 0mm 0mm,angle=0,scale=0.45]{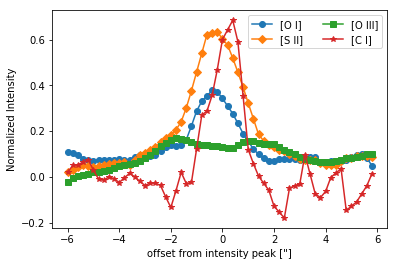} \\
  \end{array}$ 
  \caption{
    \textit{Left:} Intensity tracing through the eastern limb of the irradiated outflow.  
    \textit{Middle:} [O~{\sc i}] image showing the location of the slices through the eastern and western limbs of the jet. 
    \textit{Right:} Intensity tracing through the western limb of the irradiated outflow.
  }\label{fig:outflow_slices} 
\end{figure*}

Nebular emission lines like H$\alpha$ and [S~{\sc ii}] both trace the wide-angle irradiated outflow, as also seen in HH~666 \citep{reiter2015_hh666}. 
The MUSE data also provide multiple forbidden oxygen emission lines of different degrees of ionization.
All of the detected oxygen lines trace the wide-angle outflow (see Figure~\ref{fig:intensity_maps}). 
[O~{\sc i}] 6300~\AA\ has the same ionization potential as hydrogen (see Table~\ref{t:ionization_potentials}) and follows the same tapered shape delineated by H$\alpha$ emission from the ionized inner outflow.
[O~{\sc i}] 6300~\AA\ is not detected in the terminal bow shocks. 
[O~{\sc ii}] 7319,7330~\AA\ emission is bright throughout the system, including both bow shocks.
In contrast, [O~{\sc iii}] 5007~\AA\ is not seen in emission. 
Instead, the silhouette of the globule stands out against the bright H~{\sc ii} region and we detect weak absorption at the position of the outflow (see Figure~\ref{fig:outflow_slices}).
We see similar weak absorption of the high-excitation [S~{\sc iii}] line.

\citet{reiter2017_dust} saw a similar trend in [O~{\sc iii}] in the knots of the dusty jet HH~1019.
Absorption in the high excitation lines in HH~900 suggest that there may be some dust entrained in the outflow \citep[also suggested by morphological features seen in the \emph{HST} images;][]{smith2010}.
Finally, we note that [C~{\sc i}] 8727~\AA\ emission increases at the same location where the high excitation lines are seen in aborption (see Figures~\ref{fig:line_tracings} and \ref{fig:outflow_slices}; we discuss [C~{\sc i}] further in the next paragraph). 
Strong [O~{\sc i}] emission and evidence for dust in the outflow points to the presence of low-ionization and neutral material in the outflow itself (in addition to the jet).

\begin{figure}
\centering
\includegraphics[trim=10mm 0mm 0mm 0mm,angle=0,scale=0.625]{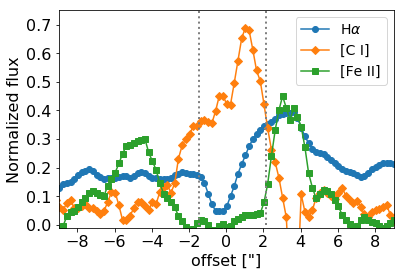} 
\caption{Emission line tracing through the HH~900 jet+outflow system showing the ionized outflow (H$\alpha$=blue), partially ionized gas ([C~{\sc i}]=orange), and predominantly neutral gas ([Fe~{\sc ii}]=green) using the same jet tracing position as shown in Figure~\ref{fig:physical_params}.  
}\label{fig:jet_lines} 
\end{figure}

We detect [C~{\sc i}] 8727~\AA\ emission from the innermost regions of the outflow (see Figures~\ref{fig:muse_intro} and \ref{fig:jet_lines}). 
The ionization potential of carbon is 11.26~eV, so the [C~{\sc i}] recombination line traces partially ionized gas and is often observed to coexist with H$_2$. 
Indeed, the [C~{\sc i}] detected with \emph{MUSE} appears to have the same morphology as the near-IR H$_2$ emission reported by \citet{reiter2015_hh900}.
The critical density of [C~{\sc i}] is $\sim 1.6 - 4.8 \times 10^4$~cm$^{-3}$ for temperatures $T=10^3 - 10^4$~K. 
We estimate densities $\sim 2 \times 10^2$~cm$^{-3}$ from the [S~{\sc ii}] lines ratio (see Section~\ref{ss:diagnostics}) from gas with similar morphology as the [C~{\sc i}] emission, suggesting that densities in the outflow are too low for significant collisional excitation.
\citet{escalante1991} argue that collisional excitation of the [C~{\sc i}] line is negligible where the fractional ionization is low, which appears to be the case in the HH~900 outflow close to the globule, where [C~{\sc i}] emission is bright (see Figure~\ref{fig:physical_params}). 

\subsection{Kinematics}\label{sss:kinematics}

\begin{figure*}
  \centering
  $\begin{array}{c}
    \includegraphics[trim=0mm 0mm 0mm 0mm,angle=0,scale=0.725]{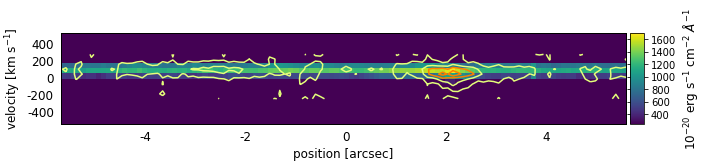} \\
    \includegraphics[trim=10mm 10mm 0mm 0mm,angle=0,scale=0.395]{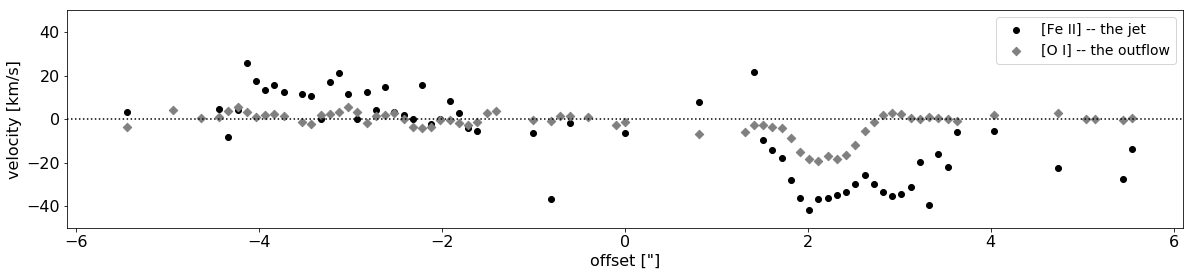} \\
\end{array}$    
  \caption{
    \textit{Top:} Position-velocity diagram of the wide-angle outflow ([O~{\sc i}] 6300~\AA; colorscale) and collimated jet ([Fe~{\sc ii}] 7155~\AA; contours) taken through the same tracing shown in Figure~\ref{fig:physical_params}. 
    \textit{Bottom:} The peak velocity at each position of the jet/outflow determined from a Gaussian fit to the line profile. 
}\label{fig:pv_diagram} 
\end{figure*}

We measure the jet velocity using [Fe~{\sc ii}] 7155~\AA\ and the kinematics of the irradiated outflow using [O~{\sc i}] 6300~\AA. 
Both lines are in the redder portion of MUSE's wavelength coverage where the velocity resolution is $\gtrsim 100$~km~s$^{-1}$. 
We fit a Gaussian to the line profiles as a function of position in the jet/outflow, allowing us to marginally resolve the velocity difference between the two lobes. 
The western limb of the jet is blueshifted compared to the eastern limb of the jet, with a velocity difference between the two limbs of $\sim 40$~km~s$^{-1}$ (see Figure~\ref{fig:pv_diagram}). 
The kinematic structure of the irradiated outflow is spectrally unresolved with MUSE, consistent with the small line-of-sight velocities reported by \citet{reiter2015_hh900}.

\section{Optical spectrum of the YSO in the western outflow limb}\label{ss:yso}
\begin{figure*}
\centering
$\begin{array}{c}
  \includegraphics[trim=0mm 0mm 0mm 0mm,angle=0,scale=0.395]{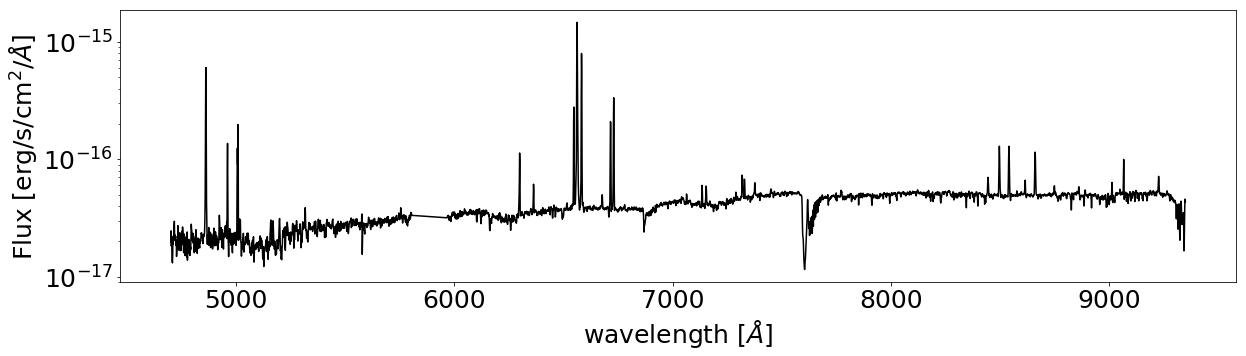} \\
  \begin{array}{cc}
    \includegraphics[trim=0mm 0mm 0mm 0mm,angle=0,scale=0.575]{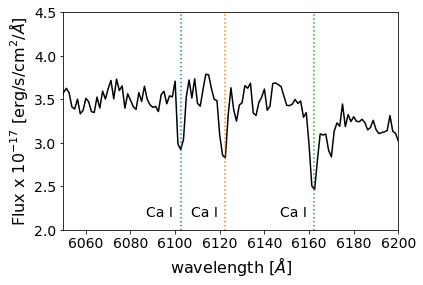} & 
    \includegraphics[trim=0mm 0mm 0mm 0mm,angle=0,scale=0.575]{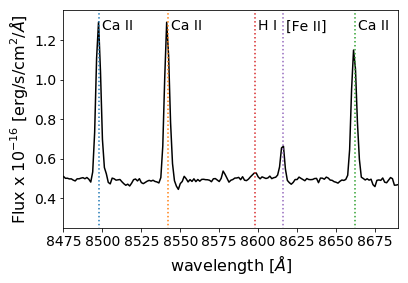} \\
  \end{array}
  \end{array}$
\caption{
  Spectrum of the YSO, PCYC~838 (see Figure~\ref{fig:muse_intro}), that lies in the western limb of the outflow (\textit{top}).
  Ca~{\sc i} absorption suggests a G or F spectral type, although model fits suggest M$_{\star} \sim 2.5$~M$_{\odot}$ (\textit{left}; see Section~\ref{ss:yso}).
  Strong Ca~{\sc ii} emission suggests ongoing accretion (\textit{right}).
}\label{fig:other_star_spec} 
\end{figure*}

The HH~900 jet-driving source is embedded in the opaque globule and cannot be seen with MUSE.
However, the less-obscured protostar to the west of the globule, identified as PCYC~838 by \citet{povich2011}, is readily detected with MUSE. 
This YSO is spatially coincident with the western limb of the HH~900 outflow and may drive a microjet of its own \citep{smith2010}. 
We perform aperture photometry to extract the spectrum of the star, although given its location, the stellar spectrum could not be cleanly separated from the outflow.
We therefore focus our analysis on lines that are unlikely to be excited in the nebular gas.

The spectrum of PCYC~838 is shown in Figure~\ref{fig:other_star_spec}. 
Continuum emission rises slightly toward redder wavelengths indicating that the source is young. 
Lines like Ca~{\sc i} $\lambda 6102,6122,6162$\AA\ 
seen in absorption suggest a G or F spectral type. 
Determining a more robust spectral type of the source is difficult given the complex structure of the nebular emission surrounding the source that prevents a clean subtraction from the stellar spectrum.
Model fits to the IR spectral energy distribution from \citet{povich2011} give 
M$_{\star} = 2.5 \pm 1.2$~M$_{\odot}$ (L$_{\star} = 2.1 \pm 1.4$~L$_{\odot}$), suggesting that the forming YSO will be an intermediate-mass star. 
In addition, excess infrared emission from the source indicates circumstellar material \citep{povich2011} 
that may feed on-going accretion which will veil the photospheric absorption lines. 
Indeed, the near-IR Ca~{\sc ii} triplet is seen in emission, which 
\citet{hillenbrand1998} showed is tightly correlated with strong accretion.
The Ca~{\sc ii} $\lambda 8542$\AA\ line is stronger than the pure H line at $\lambda 8598$\AA\ \citep[see Figure~\ref{fig:other_star_spec} and][]{hillenbrand2013}, providing additional evidence for accretion.

The spectral resolution of MUSE is inadequate to tell if emission lines likely to be excited in the putative microjet are at a different velocity than the HH~900 jet+outflow system. 
Given the clear relationship between accretion and outflow activity \citep[e.g.,][]{hartigan1995}, there may be a jet and/or outflow associated with PCYC~838 that lies close to the plane of the sky that cannot be detected at this spectral resolution.


\section{Discussion}\label{s:ir_synth_discussion}
We present new MUSE+GALACSI-AO observations of the Tadpole globule and HH~900 jet+outflow system in Carina, providing an unprecedented view of a small globule subject to intense stellar feedback.
Together with spatially-resolved maps of the cold, molecular gas (Paper~II), 
this multiwavelength view reveals the anatomy of the system and allows us to measure the external influence of the nearby high-mass stars on both the globule and the jet+outflow system.
As an isolated globule adrift in the H~{\sc ii} region, the Tadpole represents an evolutionary stage between star-forming pillars \citep[e.g.,][Klaassen et al.\ 2019, subm.]{klaassen2014}
and revealed photoevaporating protoplanetary disks \citep[e.g.,][]{odell1994}.
Understanding this intermediate stage is important for developing a more complete view of star formation in feedback-dominated environments.
In the following discussion, we consider the fate of the Tadpole globule in the context of its environment.

\subsection{Estimating local / incident radiation}\label{ss:irradiation}

Physical properties in the ionization front depend on the strength and shape of the ionizing spectrum \citep{sankrit2000}.
The Tadpole resides near Tr16 which has a well-studied population of OB stars with tabulated spectral types and ionizing photon luminosities \citep[e.g.,][]{smith2006_energy,alexander2016}.
Several morphological clues suggest that the Tadpole lies in front of Tr16:
(1) the globule has a $v_{LSR} \approx -30$~km~s$^{-1}$ compared to $v_{LSR} \approx -20$~km~s$^{-1}$ for Carina \citep[see Paper~II and][]{rebolledo2016};
(2) the Tadpole tail lies in front of the eastern outflow limb \citep{smith2010} and is blueshifted with respect to the head, as though tracing the remnants of a pillar that points away from the observer, toward the star cluster (Paper~II); 
(3) bright emission lines from the ionization front are detected on both the side of the Tadpole closest to Tr16 and on the far side of the globule (see Figure~\ref{fig:muse_intro}); and
(4) the center of the globule appears to be opaque, suggesting little illumination of the near side of the globule.

Excitation diagnostics from the MUSE data allow us to further constrain the local irradiation of the Tadpole. 
The ratio [O~{\sc iii}]/H$\alpha$ is a proxy for the spectral type of the dominant ionizing source as it provides an estimate of the ratio of photons with energy to ionize H (13.6~eV) and those adequately energetic to further ionize O$^+$ (35.1~eV, see Table~\ref{t:ionization_potentials}). 
In the Tadpole globule and HH~900 jet+outflow system, the ratio varies between $\sim 0.25-0.35$, suggesting an ionizing source with spectral type $\sim$O8 or later \citep{sankrit2000}. 
This is somewhat later than the spectral type of the nearest O-type star seen in projection, the O5~V star CPD-59~2641 \citep[see][]{reiter2015_hh900}.

In photoionization equilibrium, the density in the ionization front will increase in proportion to the overall ionizing flux \citep{sankrit2000}.
\citet{reiter2015_hh900} used this to estimate $n_{IF} \approx 4000$~cm$^{-3}$ by assuming that the nearest O-type star dominates the excitation. 
This density may be computed as 
\begin{equation}
n_{IF} = \sqrt{ \frac{Q_H}{8 \pi D^2 r \alpha_B} }
\end{equation}
where
$Q_H$ is the ionizing photon luminosity,
$r$ is the radius of curvature of the globule, and 
$D$ is the projected distance between the ionizing source and the globule. 
Ambiguities in the projected separations contribute as $\sim D^{-1}$ and thus are a larger source of uncertainty than the ionizing flux of a given star (which goes as $\sim \sqrt{Q_H}$).
This uncertainty will affect the estimated photoevaporation rate (See Section~\ref{ss:survival}), particularly when no independent density measurement is available.

Fortunately, the MUSE data allow us to make an independent measurement of the density (see Section~\ref{ss:diagnostics}).
This allows us to invert the computation and estimate the incident ionizing flux required to create the observed density. 
Assuming ionizations are balanced by recombinations in a pure hydrogen nebula, 
\begin{equation}
 \frac{Q_H}{4 \pi D^2} = a n_e n_p L \alpha_{eff}^{H\alpha}
\end{equation}\label{eq:ion_balance}
where
$Q_H$ is the luminosity of ionizing photons in $s^{-1}$, 
$D$ is the distance in pc, 
$a$ is a dimensionless factor of order 1 that accounts for geometric effects,  
$n_e = n_p$ are the electron and proton (H$^+$) density, respectively,
$L$ is the emitting path length, and 
$\alpha_{eff}^{H\alpha}$ is the recombination coefficient.  
Assuming the median distance to the OB stars in Tr16 from \citet{alexander2016} and taking
the emitting path length to be $L \approx 0.003$~pc \citep{smith2010},
we find $log(Q_H) \sim 48.3$, two orders of magnitude less than the cumulative ionizing photon luminosity from Tr16 \citep[$log(Q_H)=50.77$, see][]{smith2006_energy}. 
This suggests that the Tadpole/HH~900 are not subject to or shielded from the most extreme feedback from Tr16, despite its close projected separation.
Nevertheless, feedback is clearly affecting the jet and globule and may be rapidly clearing the remaining reservoir of gas and dust.

\subsection{The Tadpole as a pressure-bounded Bonnor-Ebert Sphere}\label{ss:mass_est}

The prototypical cloud model used in many theoretical investigations of the impact of ionizing radiation is the isothermal Bonnor-Ebert (BE) sphere \citep{bonner1956,ebert1955}.
While the model assumes equilibrium, it is unstable to collapse if sufficiently compressed. 
\citet{haworth2015} used pressure-bounded BE spheres to model the stability and evolution of small globulettes in Carina cataloged by \citet{grenman2014}, including the Tadpole.

We revisit the parameters of the Tadpole assuming it is well described by a BE sphere but using the physical parameters we measure with the MUSE data. 
The key difference in our approach compared to that of \citet{haworth2015} is that we consider the pressure provided by ongoing ionization on the globule surface, rather than assuming that the hot, diffuse gas of the H~{\sc ii} region confines the globule.
In this way, we examine the stability of the globule against collapse under the influence of external ionizing radiation, in rough analogy to the approach of \citet{bertoldi1989}.
For sufficiently low incident ionization and/or adequately high column densities, clouds will be relatively unaffected by the influence of the ionizing radiation.

Following \citet{hartmann2009}, we compute the BE mass as
\begin{equation}
  M_{BE} = 1.8 \frac{c_s^4}{G^{3/2}} \frac{1}{\sqrt{p_0}}
\end{equation}
where 
$c_s = \sqrt{k_B T / \mu m_H}$ is the sound speed in the globule, and 
$p_0$ is the gas pressure of the confining medium, $p_0 = n_e k_B T_e$. 
Assuming the temperature in the globule is $T \approx 30$~K \citep{roccatagliata2013} and a mean molecular weight of $\mu=2.3$, we find $c_s \approx 0.33$~km~s$^{-1}$ in the cold gas. 
Together with the density and temperature we measure from the MUSE data ($n_e \sim 200$~cm$^{-3}$ and $T\sim 10^4$~K, see Figure~\ref{fig:physical_params} and Section~\ref{ss:diagnostics}) to compute the confining pressure, we estimate a globule mass $M_{BE} \sim 3.7$~M$_{\odot}$. 
This is two orders of magnitude higher than the $\sim 14$~M$_{\mathrm{Jupiter}}$ (or 0.01~M$_{\odot}$) estimated by \citet{grenman2014} and regarded as a lower limit due to superimposed bright nebulosity. 
Indeed, subsequent single-dish observations of some of the Carina globulettes (not including the Tadpole) found systematically higher masses than those estimated from the extinction \citep{haikala2017}.

Using $M_{BE}$, we compute the corresponding radius of the BE sphere, $R_{BE} = 0.41 \frac{G}{c_S^2} M_{BE} \approx 12,500$~AU, or $\sim 5.4^{\prime\prime}$ at a distance of 2.3~kpc, roughly twice the observed major axis of the globule ($\sim 1.5^{\prime\prime}$). 
The BE radius corresponding to the mass estimated by \citet{grenman2014}, $\sim 0.02^{\prime\prime}$ would be unresolved, even with \emph{HST}.

Assuming a uniform density, we can use the BE mass to estimate the column density.
This allows us to compare more directly with the approach taken by \citet{grenman2014} who used $A_V$ estimates to infer the column density, and thus mass of small globulettes.
Given our higher mass, we estimate a column density and extinction $\sim 100\times$ higher than that obtained by \citet{grenman2014}. 
A 14~M$_{\mathrm{Jupiter}}$ uniform-density globule with the size of the Tadpole corresponds to a column density $N(H) \approx 3 \times 10^{21}$~cm$^{-2}$, or an $A_V \approx 1.8$~mag. 
This is within $\sim 1$~mag of the $A_V$ we estimate in the ionized gas using the ratio of hydrogen recombination lines (see Section~\ref{ss:reddening} and Figure~\ref{fig:physical_params}).
In contrast, from $M_{BE}$, we estimate $N(H) \approx 6 \times 10^{23}$~cm$^{-2}$, corresponding to $A_V \gtrsim 450$~mag.
We find the latter estimate more physically realistic given that the jet morphology and kinematics demand a protostar embedded in the opaque globule that remains unseen even in the near-IR, requiring a high extinction to obscure the driving source of one of the most powerful jets in Carina.

\citet{bertoldi1989} explored many of these same parameters to determine the fate of initially neutral clouds subject to external ionizing radiation.
The stability of clouds depends on the column density ($A_V$), and the strength of the incident radiation.
\citet{bertoldi1989} defined regions of this parameter space more or less stable against the compression from external ionization. 
For both the BE mass estimate and that produced by \citet{grenman2014}, the Tadpole parameters suggest that the cloud will be compressed by the ionization front \citep[region II in Figure~1 of][]{bertoldi1989}.
Within the \citet{bertoldi1989} framework, we expect that more ionizing photons will be absorbed before reaching the ionization front at higher column densities (the BE estimate) and the fractional size of the ionization front, compared to the size of the cloud itself, will be smaller.
Using the emitting path length of ionizing photons $L\approx0.003$~pc, we estimate the fractional size of the ionization front compared to the globule is $\sim0.3$, larger than Bertoldi estimate even for the \citet{grenman2014} mass.
Since the H$\alpha$ emission will also trace the photoevaporative flow, a more representative measure of the size of the ionization front may be the offset between emission lines of different ionization potentials \citep[e.g.,][]{carlsten2018}, but we cannot resolve the separation between the characteristic emission lines in the MUSE data even with AO (see Figure~\ref{fig:line_tracings}).

This simplistic analysis neglects some key observational features of the system, most notably that there is unambiguous evidence that the globule has already undergone collapse and formed a protostar.
The free fall time for a cloud with the Tadpole's estimated mass and radius is $<0.1$~Myr ($<0.2$~Myr for the \citealt{grenman2014} mass estimate), suggesting that the globule has already contracted from an equilibrium sphere.
In addition, the existence of the embedded jet-driving source suggests that the globule is likely not isothermal. 
More subtly, most of the assumptions that we have made presume the present-day distribution and flux of ionizing sources.
However, until recently, $\eta$~Carinae, located $\sim 2.5$~pc in projection from the Tadpole, dominated the ionizing flux of Tr16 with an estimated main-sequence ionizing photon luminosity $\log(Q_H) \approx 50.36$~s$^{-1}$.
On the main sequence, $\eta$~Car alone would increase the incident ionizing photon luminosity by an order of magnitude.

\subsection{Globule survival}\label{ss:survival}

Assuming the Tadpole is a sphere, \citet{reiter2015_hh900} estimated the photoevaporation rate as
\begin{equation} 
\dot{M}_{phot} \simeq 4 \pi r^2 \mu m_H n_H v
\end{equation}\label{eq:mdot_phot} 
where
$r \approx 0.01$~pc is the radius of curvature of the globule, 
$m_H$ is the mass of hydrogen, 
$n_H \approx 1.4 n_e \sim 10^4$~cm$^{-3}$ is the density of neutral hydrogen \citep{sankrit2000}, and  
$v \approx c_{II}$ is the speed of the evaporative flow, assumed to be the sound speed in ionized gas. 
The resulting photoevaporation rate of $\dot{M} \sim 7 \times 10^{-6}$~M$_{\odot}$~yr$^{-1}$ suggests that a 1~M$_{\odot}$ globule will survive an additional $\lesssim 10^6$~yr.

Our new MUSE data allow for better constraints on the physical properties of the globule. 
Using the density determined from the [S~{\sc ii}] line ratio (see Section~\ref{ss:diagnostics}), we find a value that is an order of magnitude lower than \citet{reiter2015_hh900} infer from the ionizing photon flux of the nearest O-type star ($\sim 3 \times 10^2$~cm$^{-3}$ compared to $\sim 4 \times 10^3$~cm$^{-3}$, respectively).
This leads to a corresponding reduction in the estimated photoevaporation rate, $\sim 5 \times 10^{-7}$~M$_{\odot}$~yr$^{-1}$, and increase in the estimated globule lifetime.

At the same time, the structure of the wide-angle outflow emerging from the Tadpole demonstrates that significant mass may be lost to jet entrainment. 
The outflow is nearly the same width as the globule itself with dusty filaments extending into the outflow from the western edge of the globule \citep{smith2010}. 
The mass-loss rate in the ionized outflow is comparable to the photoevaporation rate, $\dot{M}_{outflow} \sim 6 \times 10^{-7}$~M$_{\odot}$~yr$^{-1}$ (from \citealt{smith2010}; we note that the lower outflow velocity measured by \citealt{reiter2017} is offset by the higher density we measure with the MUSE data in Section~\ref{ss:diagnostics}). 
However, this mass-loss rate is a lower limit as it only accounts for the ionized gas and does not include the partially ionized gas traced by [C~{\sc i}] and H$_2$ or any molecules or dust in the outflow \citep[see Section~\ref{ss:jet_props} and][Paper~II]{reiter2015_hh900}.

It is difficult to constrain how much mass-loss from jet entrainment contributes to destruction of the globule, given the time-variable nature of accretion and outflow as well as the uncertainties in the entrainment efficiency. 
As a rough estimate, we compare the kinetic energy of the jet to the gravitational binding energy of the globule. 
HH~900 is one of the most energetic jets in Carina, with $E \gtrsim 1.5 \times 10^{43}$~erg \citep{reiter2017}, and likely higher given evidence for higher densities in the jet (see Section~\ref{sss:jet}).
Assuming the Tadpole is a sphere, the gravitational binding energy $U = \frac{3}{5} \frac{GM^2}{R}$ is $U \approx 6 \times 10^{43}$~erg using $M_{BE}$ estimated in Section~\ref{ss:mass_est}. 
The kinetic energy of the jet alone is within a factor of 3 of the gravitational binding energy of the globule. 
The similarity of these numbers suggests that the jet may play a significant role in the ultimate destruction of the globule.
At the same time, the survival of the globule despite the powerful jet+outflow illustrates the inefficient transfer of momentum and energy to the natal cloud.
Observations of the cold molecular gas in the globule (Paper~II) and 
more detailed models are needed to constrain the role of the jet in the ultimate destruction of the globule.

\subsection{The fate of small globules like the Tadpole}\label{ss:future} 

\citet{gahm2007,grenman2014} argue that the small globulettes seen in H~{\sc ii} regions may contribute significant numbers of planetary mass objects even if only a small number of the existing clouds collapse. 
However, \citet{haworth2015} conclude that the globulettes are stable and unlikely to collapse without an external perturbation. 
Results for the Tadpole are significantly at odds with the conclusions of \citet{grenman2014}.
We argue that the Tadpole and other globules must be of order stellar masses (rather than planetary masses) in order to prevent rapid photoablation. 
For either mass estimate, we find the external force of the ionization front on the globule surface will compress the globule.

While the Tadpole globule itself is unambiguously star-forming, its evolution and interaction with its environment are relevant for any planets that may form around the embedded jet-driving source. 
The Tadpole shares a few characteristics with some models for the formation of the Solar System. 
Fossil evidence in Solar System meteorites indicates that the Sun formed near a source of short-lived radioactive isotopes that are synthesized in the death of high-mass stars \citep[e.g.,][]{adams2010}. 
Various authors have considered the role of a nearby supernova explosion in enriching a prestellar cloud core and triggering its collapse \citep[e.g.,][]{cameron1977,boss2018}. 
The timescales required to trigger a new star formation event are at odds with the short half-lives of the essential radioactive elements \citep[e.g.,][]{parker_dale_2016}, suggesting that direct disk enrichment is still the most likely pathway.
However, disks may be rapidly destroyed under the influence of external ionization (e.g., \citealt{winter2018,nicholson2019}, although see \citealt{richert2015}), leaving little to enrich by the time nearby stars explode as supernovae.

Young objects like the Tadpole may be a prototype for a third option that is intermediate between these well-studied possibilities.
The young dynamical age of the HH~900 jet \citep[$\sim2200$~yr, see][]{reiter2015_hh900} suggests that its driving source formed after most of Tr16, where multiple post-main-sequence stars suggest an age $\gtrsim 3$~Myr \citep{walborn1995,getman2014}. 
For the estimated Tadpole photoevaporation rate of $\sim 5 \times 10^{-7}$~M$_{\odot}$~yr$^{-1}$, a $\sim 3.7$~M$_{\odot}$ globule will be completely photoevaporated in $\sim 7$~Myr. 
Including the impact of the HH~900 jet+outflow system on the globule, this remaining lifetime may be even shorter. 
This raises the possibility that the disk-bearing YSO will emerge into the H~{\sc ii} region (similar to the proplyds in Orion) during the supernova era in Carina.

The Tadpole is one of the largest objects in the \citet{grenman2014} catalog in both its physical extent and estimated mass.
It is unclear what fraction of the other globulettes may also be of order stellar mass. 
If they have indeed been compressed in the H~{\sc ii}, then high optical depths may obscure embedded protostars, as is the case in the Tadpole. 
Nevertheless, the Tadpole provides a cautionary tale that demonstrates the limits of indirect mass estimates. 
Resolved submillimeter observations provide a more direct mass estimate as well as gas kinematics to measure the dynamical impact of feedback on the gas (Paper~II). 
In the absense of such a survey of globulettes, we urge caution in considering their contribution and influence on the bottom of the IMF.

\section{Conclusions}\label{s:conclusions}

We present new AO-assisted MUSE observations of the Tadpole globule and HH~900 jet+outflow system in the Carina Nebula.
The broad spectral coverage of MUSE allows us to measure several diagnostic ratios and determine the variation in physical parameters as a function of position throughout the system. 
We measure modest extinctions ($A_V \approx 2.5$~mag) in the system using a tracer only sensitive to the ionized gas.
Densities ($n_e \approx 200$~cm$^{-3}$) and temperatures ($T_e \approx 10^4$~K) are high in both the globule and the jet+outflow system.
There is a clear increase in the excitation and ionization of the outflow with increasing distance from the opaque Tadpole globule.

Atomic emission lines also reveal the change in excitation in the outflow with increasing distance from the globule. 
We detect [C~{\sc i}] 8727~\AA, which traces partially ionized gas, in the inner outflow close to the globule. 
Evidence for absorption in high excitation lines like [O~{\sc iii}] and [S~{\sc iii}] in the outflow suggest the presence of dust. 
Together, this demonstrates that the HH~900 outflow is not fully ionized. 
The collimated jet that threads the ionized HH~900 outflow is seen in high-density tracers, suggesting that the density, and thus the mass-loss rate, in the jet may be as much as an order of magnitude higher than estimated by \citet{reiter2015_hh900}. 
As a result, the kinetic energy of the jet alone (not including the outflow) is within a factor of three of the gravitational binding energy of the Tadpole.

We extract the spectrum of the YSO that lies projected onto the western limb of the HH~900 outflow.
While contamination from the environment prevents robust spectral typing, bright emission from the near-IR Ca~{\sc ii} triplet indicates strong on-going accretion, consistent with evidence for a circumstellar disk presented by \citet{povich2011}. 
The protostar that drives the HH~900 jet+outflow system is embedded within the opaque Tadpole globule and not detected with MUSE.

We 
estimate a Bonnor-Ebert mass of $\sim 3.7$~M$_{\odot}$ for the globule, $\sim 100\times$ larger than previous estimates. 
Assuming a uniform density globule, we estimate a column density $N(H) \sim 6 \times 10^{23}$~cm$^{-2}$, corresponding to $A_V \gtrsim 450$~mag, in agreement with a globule so opaque that the HH~900 driving source remains unseen even in the infrared. 
The electron density in the ionization front is lower than previous estimates, leading to a smaller photoevaporation rate and longer survival time of the globule assuming that photoevaporation dominates the destruction of the globule.
Observations of the cold, molecular gas in the globule (Paper~II) will allow us to test these estimates.

The results from this MUSE study suggest that small globules may play an interesting role in the H~{\sc ii} region ecosystem. 
Stars that form in small globules will be exposed to the most intense feedback later in the life of the nearby high-mass stars.
This may provide a pathway for planet-forming disks to survive destructive ionizing radiation and thus be enriched with essential elements synthesized in the death of high-mass stars. 
While we present strong evidence that the Tadpole globule is orders of magnitude more massive than previous estimates, it is not clear what fraction of the globulettes may be stellar-mass objects (rather than planetary-mass).
Nevertheless, the discrepancy in the Tadpole mass estimates point to the need for caution when considering the contribution that globulettes might make to the bottom of the IMF.


\section*{Acknowledgements}
We thank the referee, Dr G\"{o}sta Gahm, for a timely and thoughtful report. 
MR would like to thank Libby Jones for helpful discussions and Tom Haworth for a careful reading of the manuscript. 
In loving memory of John Causland. 
This project has received funding from the European Union's Horizon 2020 research and innovation programme under the Marie Sk\'{l}odoska-Curie grant agreement No. 665593 awarded to the Science and Technology Facilities Council. 
JCM acknowledges
support from the European Research Council under the European Community’s
Horizon 2020 framework program (2014-2020) via the ERC Consolidator
grant ‘From Cloud to Star Formation (CSF)’ (project number 648505).
This paper is based on data obtained with ESO telescopes at the Paranal Observatory under programme ID 0101.C-0391(A). 
This research made use of Astropy,\footnote{http://www.astropy.org} a community-developed core Python package for Astronomy \citep{astropy:2013, astropy:2018};
APLpy, an open-source plotting package for Python \citep{robitaille2012}; and
PySpecKit \citep{ginsburg2011}.




\bibliographystyle{mnras}
\bibliography{../../../bibliography_mrr}



\appendix

\section{Lines detected in the Tadpole}

\begin{onecolumn}
\begin{center}
\begin{footnotesize}
\begin{longtable}{lll}
\caption{Lines detected in MUSE data}\label{t:MUSE_lines} \\ 
\hline\hline
Line & $\lambda$ & component \\ 
name & [\AA] &  \\
\endfirsthead
\hline\hline
H$\beta$ & 4861.49 & jet \\
$[$O~{\sc iii}$]$ & 4958.92 & globule silhouette \\ 
$[$O~{\sc iii}$]$ & 5006.84 & globule silhouette \\ 
$[$N~{\sc ii}$]$ & 5754.64 & outflow, globule edge \\
$[$O~{\sc i}$]$ & 6300.30 & outflow \\
$[$S~{\sc iii}$]$ & 6312.10 & bow shocks, outflow \\
$[$O~{\sc i}$]$ & 6363.78 & outflow \\
$[$N~{\sc ii}$]$ & 6548.03 & outflow, bow shocks, globule edge \\
H$\alpha$ & 6562.82 & outflow, bow shocks, globule edge \\
C~{\sc ii}? & 6578.03 & haze around globule \\
$[$N~{\sc ii}$]$ & 6583.41 & outflow, bow shocks, globule edge \\
He~{\sc i} & 6678.15 & faint outflow, bow shocks \\
$[$S~{\sc ii}$]$ & 6716.47 & outflow, bow shocks, globule edge \\
$[$S~{\sc ii}$]$ & 6730.85 & outflow, bow shocks, globule edge \\ 
He~{\sc i} & 7065.28 & outflow, bow shocks, globule edge \\
$[$Ar~{\sc iii}$]$ & 7135.78 & outflow, bow shocks, globule edge \\
$[$Fe~{\sc ii}$]$ & 7155.14 & jet \\
He~{\sc i} & 7281.35 & outflow, bow shocks, globule edge \\
$[$Ca~{\sc ii}$]^*$ & 7291.47 & jet \\
$[$O~{\sc ii}$]$ & 7319.65 & outflow, bow shocks, globule edge \\
$[$Ca~{\sc ii}$]^*$ & 7323.89 & jet \\
$[$O~{\sc ii}$]$ & 7330.16 & outflow, bow shocks, globule edge \\
$[$Ni~{\sc ii}$]$ & 7377.83 & jet \\
$[$Fe~{\sc ii}$]$ & 7452.50 & jet \\
$[$Fe~{\sc ii}$]^*$ & 7637.54 & jet \\
$[$Ar~{\sc iii}$]$ & 7751.12 & outflow, globule edge \\
Paschen 24-3 & 8333.78 & globule edge, brightest in inner outflow \\
Paschen 22-3 & 8359.01 & inner outflow, globule edge \\ 
Paschen 21-3 & 8374.48 & inner outflow, globule edge \\ 
Paschen 20-3 & 8392.42 & inner outflow, globule edge \\ 
 Paschen 19-3 & 8413.32 & inner outflow, globule edge \\ 
Paschen 18-3 & 8437.96 & inner outflow, globule edge \\ 
O~{\sc i} & 8446.48 & outflow \\
Paschen 16-3 & 8502.49 & inner outflow, globule edge \\
Paschen 15-3 & 8545.38 & outflow, globule edge \\
$[$Cl~{\sc ii}$]$ & 8578.70 & outflow, bow shocks, globule edge \\
Paschen 14-3 & 8598.39 & outflow, globule edge \\
$[$Fe~{\sc ii}$]$ & 8616.96 & jet \\
Paschen 13-3 & 8665.02 & outflow, bow shocks, globule edge  \\
$[$C~{\sc i}$]^{**}$ & 8727.12 & globule edge, H$_2$-bright inner outflow \\
Paschen 12-3 & 8750.48 & outflow, globule edge \\
Paschen 11-3 & 8862.79 & outflow, globule edge, bow shocks \\
$[$Fe~{\sc ii}$]$ & 8891.88 & jet \\
Paschen 10-3 & 9014.91 & ouflow, globule edge, bow shocks \\
$[$Fe~{\sc ii}$]^*$ & 9051.95 & jet \\
$[$S~{\sc iii}$]$ & 9068.90 & outflow, globule edge, bow shocks \\
Paschen 9-3 & 9229.02 & outflow, globule edge, bow shocks \\
\hline 
\multicolumn{3}{l}{$^*$ ID from \citet{ell13}; $^{**}$ ID from \citet{liu1995}} \\
\end{longtable}
\end{footnotesize}
\end{center}
\end{onecolumn}

\section{Intensity maps}

\begin{figure*}
  \centering
$\begin{array}{cc}
    \includegraphics[trim=25mm 15mm 0mm 10mm,angle=0,scale=0.365]{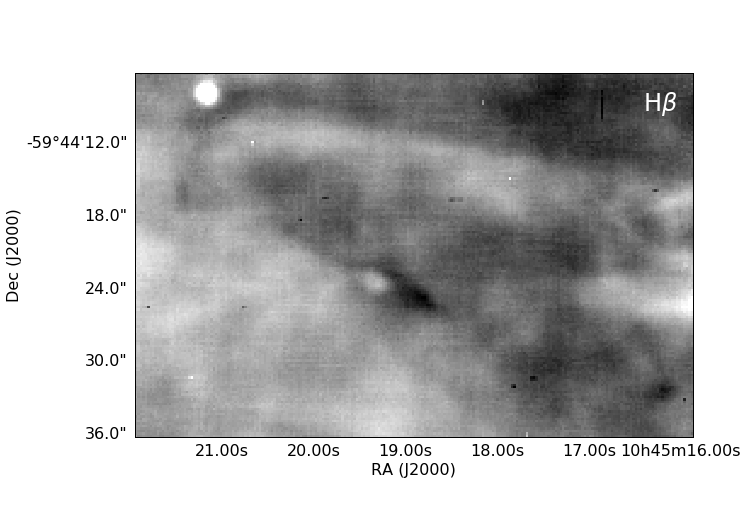}  &
    \includegraphics[trim=20mm 15mm 0mm 10mm,angle=0,scale=0.375]{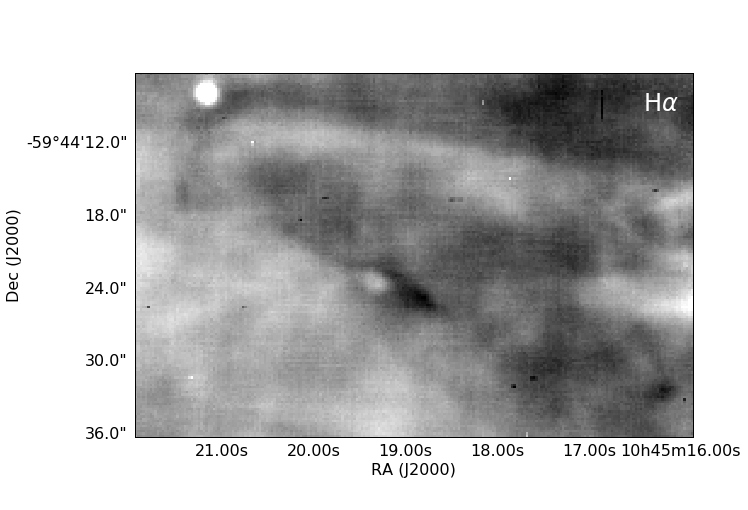} \\

    \includegraphics[trim=25mm 15mm 0mm 10mm,angle=0,scale=0.365]{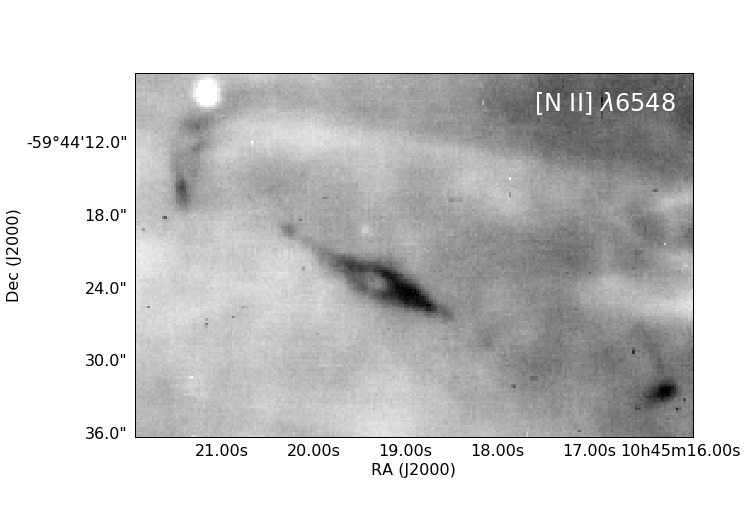}  &
    \includegraphics[trim=20mm 15mm 0mm 10mm,angle=0,scale=0.375]{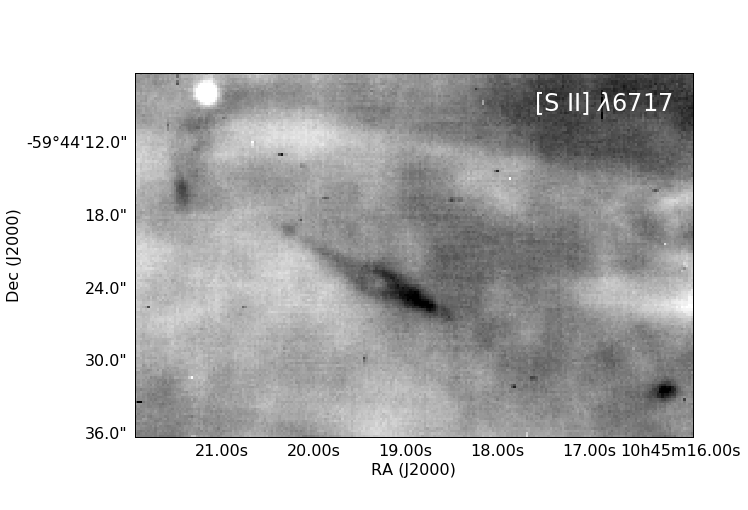} \\

    \includegraphics[trim=25mm 15mm 0mm 10mm,angle=0,scale=0.365]{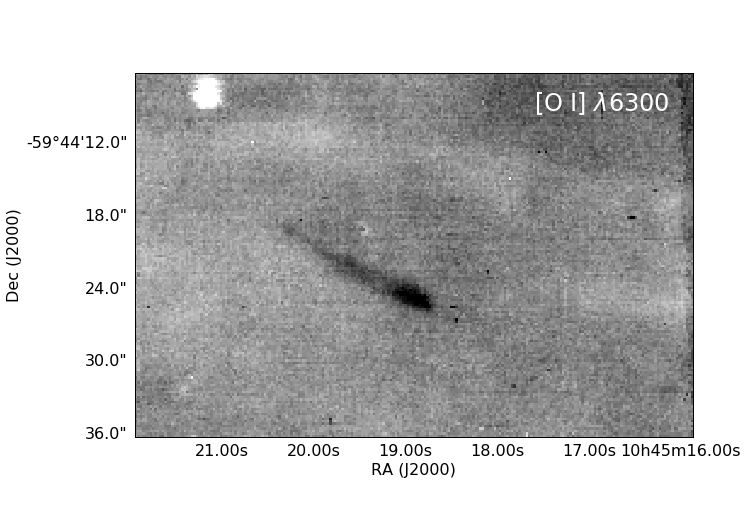}  &
    \includegraphics[trim=20mm 15mm 0mm 10mm,angle=0,scale=0.375]{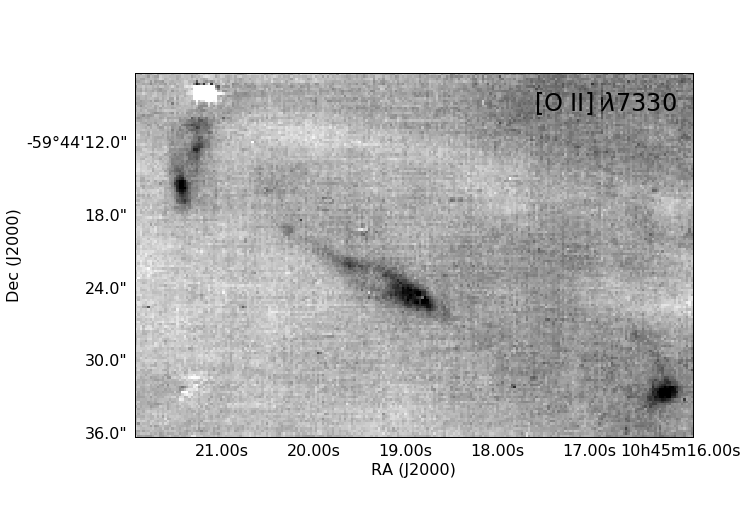} \\

    \includegraphics[trim=25mm 15mm 0mm 10mm,angle=0,scale=0.365]{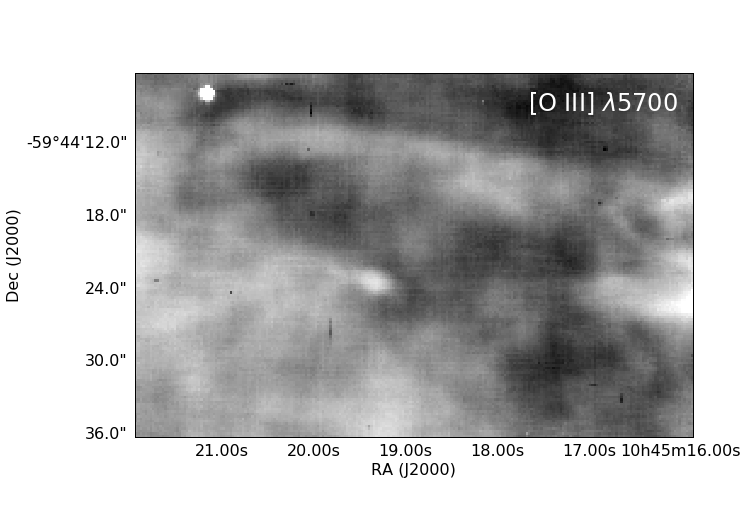}  &
    \includegraphics[trim=20mm 15mm 0mm 10mm,angle=0,scale=0.375]{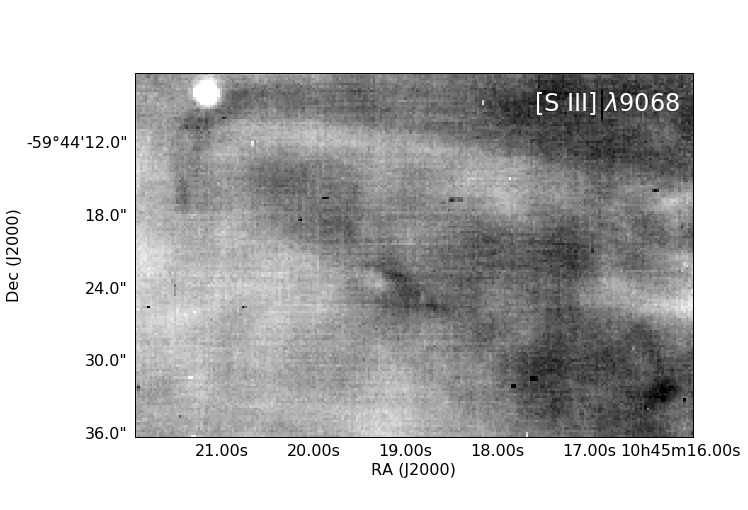} \\
  \end{array}$
  \caption{Continuum-subtracted integrated intensity maps of the Tadpole globule and the HH~900 jet+outflow system.
  }\label{fig:intensity_maps} 
\end{figure*}

\begin{figure*}
  \centering
$\begin{array}{cc}
    \includegraphics[trim=25mm 15mm 0mm 10mm,angle=0,scale=0.365]{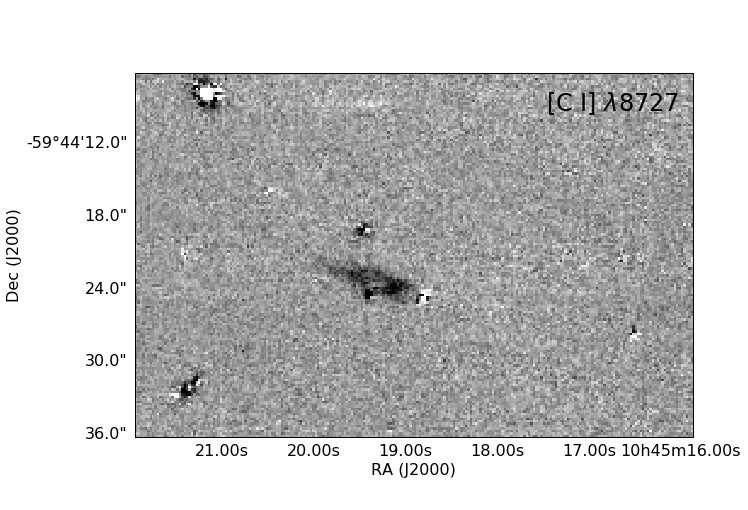}  &
    \includegraphics[trim=20mm 15mm 0mm 10mm,angle=0,scale=0.375]{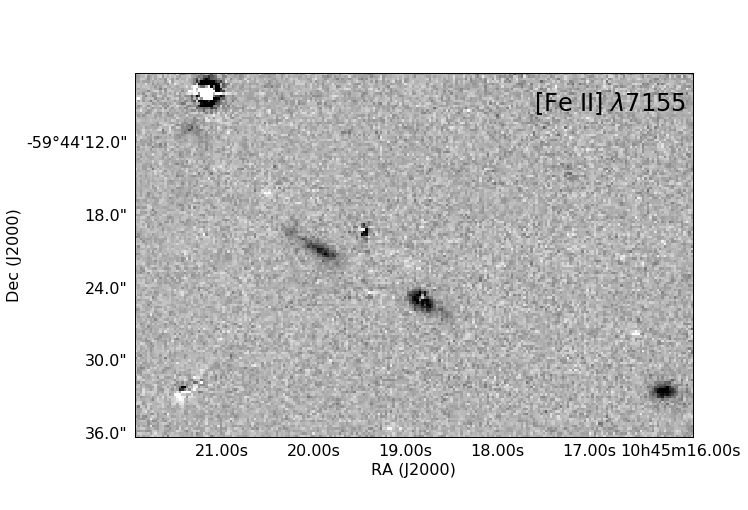} \\    
  \end{array}$
  \caption{(continued) Continuum-subtracted integrated intensity maps of the Tadpole globule and the HH~900 jet+outflow system.
  }\label{fig:intensity_maps_p2} 
\end{figure*}

\clearpage

\section{Line ratio maps}

\begin{figure*}
  \centering
$\begin{array}{cc}
    \includegraphics[trim=25mm 15mm 0mm 10mm,angle=0,scale=0.365]{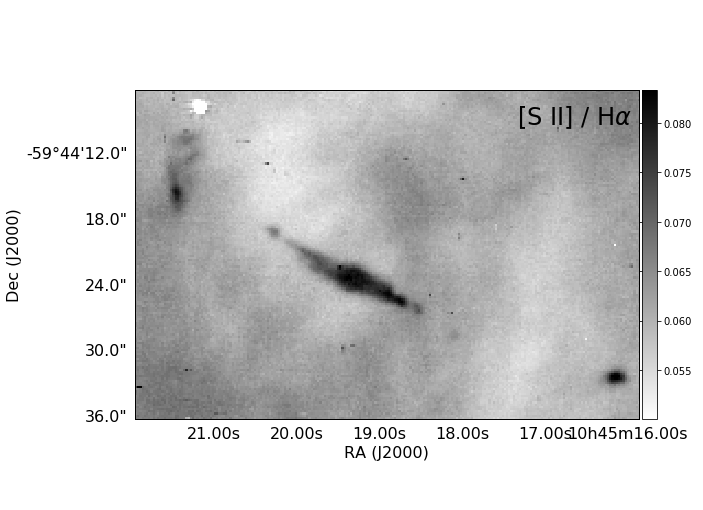}  &
    \includegraphics[trim=20mm 15mm 0mm 10mm,angle=0,scale=0.375]{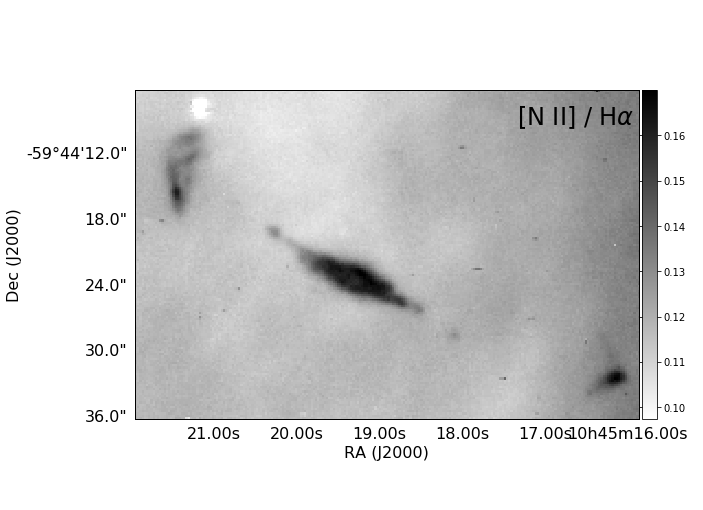} \\    
    \includegraphics[trim=25mm 15mm 0mm 10mm,angle=0,scale=0.365]{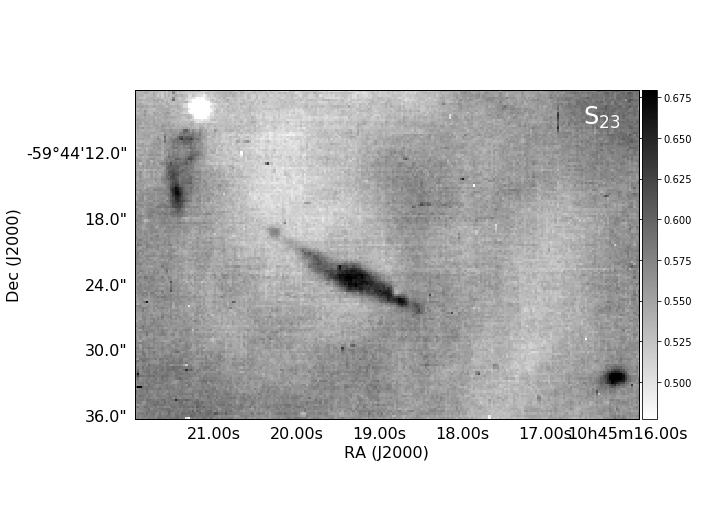}  &
    \includegraphics[trim=20mm 15mm 0mm 10mm,angle=0,scale=0.375]{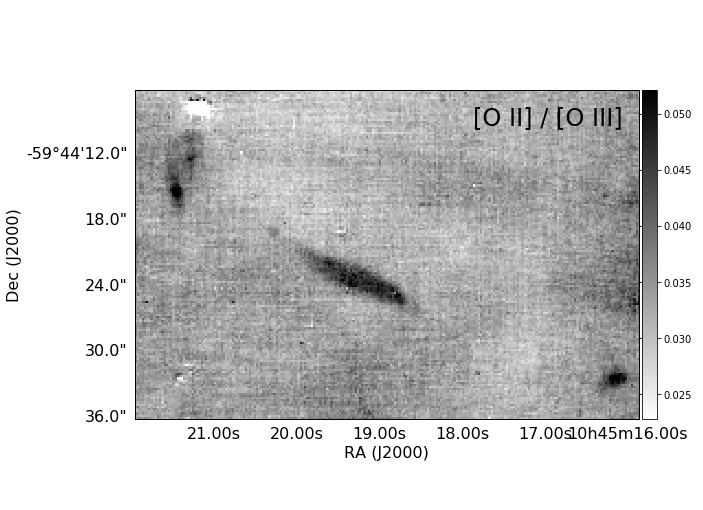} \\    
  \end{array}$
  \caption{Line ratio maps of the Tadpole globule and the HH~900 jet+outflow system.
  }\label{fig:ratio_maps} 
\end{figure*}


\bsp	
\label{lastpage}
\end{document}